\documentclass[sn-mathphys,Numbered]{sn-jnl}

\usepackage{graphicx}%
\usepackage{multirow}%
\usepackage{amsmath,amssymb,amsfonts}%
\usepackage{amsthm}%
\usepackage{mathrsfs}%
\usepackage[title]{appendix}%
\usepackage{xcolor}%
\usepackage{textcomp}%
\usepackage{manyfoot}%
\usepackage{booktabs}%
\usepackage{algorithm}%
\usepackage{algorithmicx}%
\usepackage{algpseudocode}%
\usepackage{listings}%
\usepackage{caption}%

\theoremstyle{thmstyleone}%
\theoremstyle{thmstyletwo}%

\theoremstyle{thmstylethree}%

\begin{document}

\title[Article Title]{\textbf{Programming hydrogel adhesion with engineered polymer network topology}}


\author[1]{\fnm{Zhen} \sur{Yang}}\email{zhen.yang3@mail.mcgill.ca}

\author[1]{\fnm{Guangyu} \sur{Bao}}\email{guangyu.bao@mail.mcgill.ca}

\author[1]{\fnm{Shuaibing} \sur{Jiang}}\email{shuaibing.jiang@mail.mcgill.ca}

\author[2]{\fnm{Xingwei} \sur{Yang}}\email{xingwei.yang@colorado.edu}

\author[1]{\fnm{Ran} \sur{Huo}}\email{ran.huo3@mail.mcgill.ca}

\author[1]{\fnm{Xiang} \sur{Ni}}\email{xiang.ni2@mail.mcgill.ca}

\author[1,3]{\fnm{Luc} \sur{Mongeau}}\email{luc.mongeau@mcgill.ca}

\author[2]{\fnm{Rong} \sur{Long}}\email{rong.long@colorado.edu}

\author*[1,3]{\fnm{Jianyu} \sur{Li}}\email{jianyu.li@mcgill.ca}

\affil[1]{\orgdiv{Mechanical Engineering}, \orgname{McGill University}, \orgaddress{\city{Montreal}, \postcode{H3A 0C3}, \state{QC}, \country{Canada}}}

\affil[2]{\orgdiv{Mechanical Engineering}, \orgname{Colorado University Boulder}, \orgaddress{\city{Boulder}, \postcode{80309}, \state{CO}, \country{USA}}}

\affil[3]{\orgdiv{Biomedical Engineering}, \orgname{McGill University}, \orgaddress{\city{Montreal}, \postcode{H3A 2B4}, \state{QC}, \country{Canada}}}


\abstract{Hydrogel adhesion that can be easily modulated in magnitude, space, and time is desirable in many emerging applications ranging from tissue engineering, and soft robotics, to wearable devices. In synthetic materials, these complex adhesion behaviors are often achieved individually with mechanisms and apparatus that are difficult to integrate. Here, we report a universal strategy to embody multifaceted adhesion programmability in synthetic hydrogels. By designing the surface network topology of a hydrogel, supramolecular linkages that result in contrasting adhesion behaviors are formed on the hydrogel interface. The incorporation of different topological linkages leads to dynamically tunable adhesion with high-resolution spatial programmability without alteration of bulk mechanics and chemistry. Further, the association of linkages enables stable and tunable adhesion kinetics that can be tailored to suit different applications. We rationalize the physics of chain slippage, rupture, and diffusion that underpins emergent programmable behaviors. We then incorporate the strategy into the designs of various devices such as smart wound patches, fluidic channels, drug-eluting devices, and reconfigurable soft robotics. Our study presents a simple and robust platform in which adhesion controllability in multiple aspects can be easily integrated into a single design of a hydrogel network.}




\maketitle

\newpage
\section*{Introduction}\label{sec1}

The ability to program hydrogel adhesion has significant implications in engineering, biology, and medicine. The variables of hydrogel adhesion include adhesion energy, spatial distribution, and kinetics. Among them, controlling adhesion energy is needed for bonding reinforcement after placement\cite{yang2021stimulation} or for easy detachment without damaging the adherend surface\cite{cho2019intrinsically,croll2019switchable}. Controlling the spatial distribution enables independent modulation of adhesion at different locations and can be useful for wound dressings that require strong attachment to healthy tissues while preventing stickiness to fragile and delicate wound beds. While most existing research efforts focused on the adhesion magnitude at the equilibrium stage, controlling adhesion kinetics, i.e., to modulate transient adhesion temporally, allows one to tune the operating time window for adhesive placement and is equally important but less explored. Programming the multifaceted adhesion with high-level control could enable and improve various applications ranging from tissue repair to soft robotics, yet remains extremely challenging.  

In nature, marine animals such as \textit{flatworms} control adhesion to substrates using sophisticated adhesion organs that contain two glands which respectively release adhesive and de-adhesive agents\cite{lengerer2018properties}. Such programmable adhesion is difficult to achieve for synthetic adhesives because they require the addition of complex chemistry and apparatus that are potentially difficult to integrate. For instance, adhesion based on covalent bonds is generally strong but difficult to modulate \cite{Yuk2016} unless introducing specific chemistries \cite{chen2020instant}. Physical interactions offer more flexibility to modulate adhesion energy, but require specific material properties (e.g., viscoelasticity) or additional apparatus (light, ultrasound, etc.) \cite{Yang2018,gao2019photodetachable,ma2022controlled}. In terms of adhesion kinetics, the rate of covalent bonding is fundamentally dependent on the specific chemical reactions involved. Although using different bonds with varying reaction kinetics can in principle enable tunable adhesion kinetics, incorporation of multiple reactions into one system is challenging. Physical interactions often form instantaneously and hence do not offer sufficient tunability in the kinetics for applications that might desire rather slow kinetics\cite{wang2019instant}. Achieving spatial control of adhesion requires forming (or suppressing) the interactions at selective locations through sophisticated surface patterning, while the outcomes could be compromised by uncontrolled diffusion of chemical reagents. 
A universal design strategy that inherently allows for robust and multifaceted adhesion programming on diverse surfaces is still missing.

\begin{figure}[h!] 
    \includegraphics[width=\textwidth]{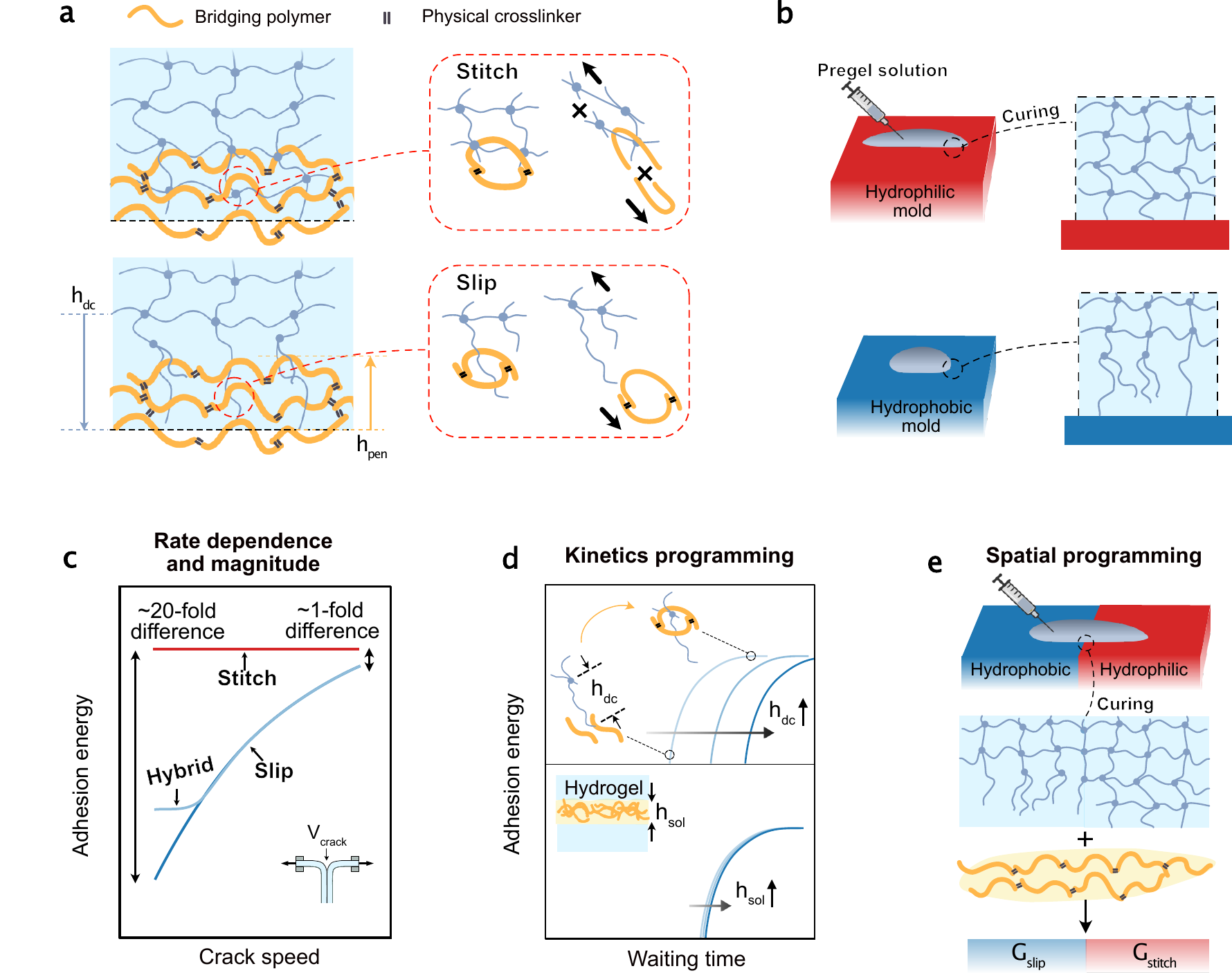}
    \caption{\textbf{Engineered network topology and linkages for multifaceted programming of hydrogel adhesion.}
    (a) Schematics of the stitch linkage (Top) and slip linkage (Bottom) formed between a bridging polymer and networks without and with surface dangling chains. The thickness of the dangling chain layer and the penetration depth of the bridging polymer are denoted as $h_{\mathrm{dc}}$ and $h_{\mathrm{pen}}$, respectively. (b) Hydrophilic and hydrophobic molds are used to form a regular network (Top) and a network carrying surface dangling chains (Bottom), respectively. (c) Rate dependence and magnitude of the adhesion energy depend on the interfacial linkage types: stitch linkages with $h_{\mathrm{pen}}/h_{\mathrm{dc}}\rightarrow \infty$, slip linkages with $h_{\mathrm{pen}}/h_{\mathrm{dc}}\ll 1$, and their hybrid with $h_{\mathrm{pen}}/h_{\mathrm{dc}}\approx 1$. (d) The slip linkage offers programmable adhesion kinetics through tuning $h_{\mathrm{dc}}$ (Top), which is also insensitive to processing conditions such as the thickness of bridging polymer solution $h_{\mathrm{sol}}$ (Bottom). (e) Spatially controllable adhesion obtained from patterning the topological linkages at the interface.
    }
    \label{fig1} 
    \end{figure}

Here we report that engineering surface network topology provides a facile, predictive, and robust methodology to program multifaceted hydrogel adhesion. This approach leverages the principles of polymer entanglement that are applicable to a wide range of materials. The underlying strategy is through two adhesion units constructed by interfacial polymer entanglements of different topologies, referred to as the slip and stitch linkages (Fig \ref{fig1}a). The slip linkage has the topology of a long polymer chain entangled with another crosslinked network (chain to network). Such a linkage can dissociate in a rate-dependent manner so that the adhesion energy can be varied over many folds without background dissipation by varying loading rate (Fig \ref{fig1}c). The association kinetics of the slip linkage dominates over other operating condition-dependent sub-kinetics and is controllable through tuning its governing length scale, enabling a stable kinetic time that can be tuned in a wide range from $\sim$50s to $\sim$1000s (Fig \ref{fig1}d). In contrast, the stitch linkage has the topology of two crosslinked polymer networks entangled together (network to network), as found in the hydrogel topological adhesion and offers less tunability in terms of adhesion energy and kinetics\cite{Yang2018}. Through a simple fabrication technique, we pattern the slip and stitch linkages spatially at the same interface (Fig 1b and 1e). Their contrasting adhesion behaviors enable pre-defined and spatially varying adhesion in a scalable manner. As such, we can embody adhesion programmability in multiple aspects in a hydrogel adhesive through a single design of the network structure, which we refer to as the topologically engineered adhesives (TEA) in this paper. The robust, facile, and predictive strategy for unprecedented control over hydrogel adhesion opens up numerous opportunities in engineering and medicine. We demonstrate the applications of our strategy in a wide range of devices, including wound patches, drug-eluting depots, fluidic channels, and soft actuators.  

\section*{Design of the interfacial topological linkages} 
To robustly program the adhesion between hydrogel adhesives and targeted surfaces, we create a diffusive interface by placing a third species of diffusive polymer, called bridging polymer, to the interface\cite{Li2017,Yang2018}. Formation of the chain-to-network topology of the slip linkage demands the following conditions: (1) the hydrogel network needs to contain dangling chains and (2) a thermodynamic driving force is needed to facilitate the diffusion of bridging polymers into the gel network. Meanwhile, the diffusion needs to be halted once the linkage forms to prevent the over-diffusion of bridging polymers into the bulk gel, which may reduce the number of linkages at the interface. 

To meet the first condition: we choose polyacrylamide (PAAm) as a model hydrogel network and polymerize it on mold with low surface tension such as Poly(methyl methacrylate) (PMMA). The hydrophobicity and other associated effects inhibit the free-radical polymerization of the gel in the vicinity of the mold \cite{Gong2001,kii2001heterogeneous,mandal2021oxygen}. This results in a surface layer of branched dangling chains with thickness $h_{\mathrm{dc}} \approx 10\sim 100 \mu$m, “protruding” from the crosslinked bulk network. In contrast, gels polymerized on molds with high surface tension such as glass are not subject to the hydrophobic mold effect, and hence contain crosslinked networks instead of branched dangling chains on their surfaces. The gels with and without engineered surface dangling chains are hereafter referred to as the TEA and regular gels, respectively. To meet the second criterion, stimuli-responsive polymers such as chitosan or gelatin were chosen as bridging polymers. The polarity of the hydrogel network and chitosan chains and the entropy of mixing promote the diffusion of chitosan chains into the hydrogel; meanwhile, the chitosan chains can be triggered to crosslink into a bridging network through a reaction-diffusion process in responding to pH changes, leading to penetration depths $h_{\mathrm{pen}}$ on the order of tens of microns\cite{Li2017}. The network formed in-situ provides a more efficient way to engage the dangling chains as opposed to a pre-formed network in which the dangling chains have to diffuse slowly through reptation to form entanglement. Other strategies to form the chain-to-network topology of slip linkage at soft material interfaces can be found in Supplementary note \ref{sec:note3}. 

Additionally, to encode adhesion kinetics and ensure the repeatable formation of slip linkages, the dominating kinetic mechanisms for the association of slip linkage should intrinsically rely on the gel network rather than being sensitive to external processing conditions that are difficult to control such as the thickness of cast solution $h_{\mathrm{sol}}$. The formation of the slip linkage is associated with the diffusion of the bridging polymers (kinetic time $t_{\mathrm{d}}$) and their gelation process (kinetic time $t_{\mathrm{gel}}$), which depend on different governing length scales. Specifically, $t_{\mathrm{d}}$ and $t_{\mathrm{gel}}$ are associated with the diffusion of bridging polymers and gelling triggers over the thicknesses of the dangling chain layer $h_{\mathrm{dc}}$ and that of the cast solution $h_{\mathrm{sol}}$, respectively. Since $h_{\mathrm{dc}}$ is a well-defined material property while $h_{\mathrm{sol}}$ is sensitive to various processing conditions\cite{Yang2018,steck2020topological}, a well-defined and controllable adhesion kinetics ensues if $t_\mathrm{slip}\geq t_\mathrm{gel}$. A simple scaling analysis allows us to determine a rough criterion to fulfill this requirement: $h_{\mathrm{dc}}^2 D_{\mathrm{eff, gel}} / h_\mathrm{sol}^2 D_{\mathrm{eff}} \geq 1$, where $D_{\mathrm{eff,gel}}$ and $D_{\mathrm{eff}}$ are the effective diffusion coefficients of gelling triggers and bridging polymers, respectively. Taking $D_{\mathrm{eff}} \approx 5\cdot 10^{-12}\mathrm{m}^{2}\mathrm{s}^{-1}$(Supplementary note \ref{sec:note3} ) and $D_{\mathrm{eff,gel}} \approx 10^{-11}\mathrm{m}^{2}\mathrm{s}^{-1}$\cite{steck2020topological} leads to $h_{\mathrm{dc}}/ h_{\mathrm{sol}} \gtrapprox 1$. $h_{\mathrm{sol}}$ at a hydrogel interface is often in the range of 10$\sim$100 $\mu$m \cite{steck2020topological,Yang2018}, which means that $h_{\mathrm{dc}}$ needs to be in the comparable range and is readily satisfied by our fabrication techniques. 

We hypothesize that the condition $h_{\mathrm{dc}}/ h_{\mathrm{sol}} \gtrapprox 1$ can lead to well-defined association kinetics of the slip linkage that encodes the overall adhesion kinetics. Once the slip linkage is formed, the long PAAm chains entangle with the crosslinked bridging network. This means that the linkage can dissociate via chain slippage, which is expected to be a thermally activated process that results in rate-sensitive adhesion as generally seen at the cell and elastomeric interfaces,  and other bonds\cite{evans1997dynamic,bell1978models,chaudhury1999rate,pobelov2017dynamic}. Unlike slip linkage, stitch linkages form when the bridging polymer diffuses into a regular gel and crosslinks into the bridging network, and their failures must involve the breaking of one of the networks, thereby leading to strong and rate-insensitive adhesion\cite{Yang2018}. 

\subsection*{Structural characterization}

\begin{figure}[h!] 
    \centering
    \includegraphics[width=0.9\textwidth]{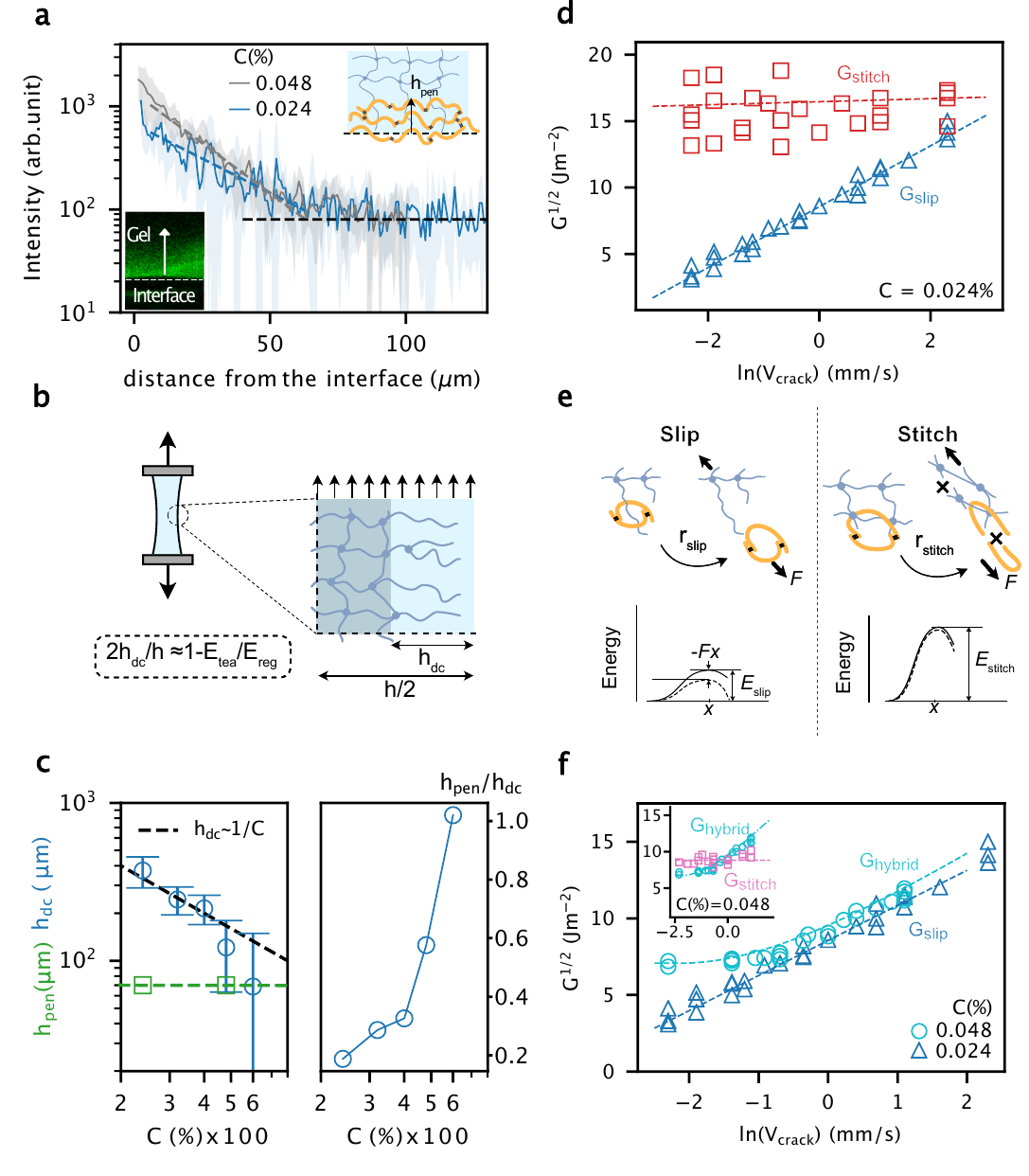}
        \caption{\textbf{Design and characterization of topology-engineered adhesive (TEA).} (a) Intensities of florescent chitosan chains diffused in TEA gels. The shaded area represents the standard deviation from 5 measures. The stronger chitosan intensity for higher $C$ may be due to more chitosan chains trapped by the denser dangling chains on the interface. (b) An idealized model used to estimate the thickness of branched dangling chain layer. (c) Left panel: estimated $h_{\mathrm{dc}}$ and measured $h_{\mathrm{pen}}$  as functions of the crosslinker-to-monomer ratio $C$. Error bars represent standard deviation. Right panel: relative engagement length $h_{\mathrm{pen}}/h_{\mathrm{dc}}$ as a function of $C$. \textbf{Dissimilar topological linkages lead to contrasting adhesion behaviors.} (d) Slip and stitch linkages-mediated $G^{1/2}$ plotted as functions of $\ln (V_{\mathrm{crack}})$ for $C=0.024\%$ and $c_{chi}=2\%$ g/mL. (e) Illustration showing the dissociations of slip and stitch linkages as thermally activated processes, with reaction rate of dissociation $r_i$ ($i$ can be slip or stitch). Upon a separation force $F$, the activation energy of the linkage is decreased by $-Fx$, where $x$ is the seperation distance. (f) The formation of topological linkages and the resulting adhesion depend on $h_{\mathrm{pen}}/h_{\mathrm{dc}}$, which is controlled by $C$. Slip and hybrid linkages are achieved for TEA gels with $C=0.024\%$ and $0.048\%$, respectively. The inset shows the same curves as (d) but for $C=0.048\%$.
        }
    \label{fig2} 
\end{figure}

Based on the above principles, we fabricate a model TEA using single-network TEA gel made of PAAm and use chitosan as the bridging polymer. 
To probe the engagement length between the dangling chains and the bridging polymer chains, we used confocal microscopy to visualize how fluorescently labelled chitosan chains penetrate the TEA gel at equilibrium. The fluorescence intensities exponentially decrease from the outermost surface to the bulk of the TEA gels with different crosslinker-to-monomer ratios $C$ (colored dash lines, Fig \ref{fig2}a). For different $C$, we measured similar distances where the intensities meet the lower plateau (black dash line, Fig \ref{fig2}a), defining the penetration depth of the bridging polymer $h_{\mathrm{pen}}\approx 70 \mu m$. This value depends on the reaction-diffusion process and thus may vary with the type of bridging polymers. For instance, $h_{\mathrm{pen}}$ for gelatin is expected to be temperature-dependent. 

As directly imaging the dangling chains is challenging, we made a first-order estimation of the thickness $h_{\mathrm{dc}}$ of the dangling chain layer from the experimentally measured elastic moduli. The TEA gel has a total thickness of $h$ and is idealized with a tri-layer model (Fig \ref{fig2}b): a layer of a regular network is sandwiched by two layers of branched dangling chains. The elastic modulus of the sandwiched regular network $E_{\mathrm{reg}}$ can be measured from a regular hydrogel formed at the same conditions except using a hydrophilic mold, given their observed structural similarity\cite{kii2001heterogeneous,mandal2021oxygen,zhang2019creating}. The elastic modulus of the dangling chain layer is assumed to be negligible since it cannot carry any transverse loads. As such, we can estimate $h_{\mathrm{dc}}$ from the ratio of measured elastic moduli of the TEA and regular gels $E_{\mathrm{tea}}/E_{\mathrm{reg}}$ in uniaxial tensile tests (Eqn. \ref{eqn:hdc} and Fig \ref{figS1}). The estimations of $h_{\mathrm{dc}}$ show a decreasing trend with the increasing value of $C$ (Fig \ref{fig2}c). The trend $h_{\mathrm{dc}} \sim C^{-1}$ may be attributed to the competition between bulk elasticity of the gel network and interface tension during gelation on hydrophobic mold\cite{Gong2001} (Supplementary Note \ref{sec:note1}), demonstrating a controlled method for fabricating the dangling chain layer of different sizes.

With the measured length scales, we calculate their ratio $h_{\mathrm{pen}}/h_{\mathrm{dc}}$ to quantify the extent to which the bridging polymers engage the dangling chains, which is expected to govern the formation of different topological linkages at the TEA gel interface (Fig \ref{figS2}d). When $h_{\mathrm{pen}}/h_{\mathrm{dc}}\ll1$, the bridging network only engages a part of the dangling chain layer, so that the interface only comprises slip linkage. If $h_{\mathrm{pen}}/h_{\mathrm{dc}} \approx 1$, a complete engagement ensues which indicates that part of the bridging polymers may diffuse across the dangling chain layer to stitch the underlying network of the TEA gel. In this case, the linkage is expected to behave as the combination of the slip and stitch linkage and is referred to as the hybrid linkage (Fig \ref{fig1}c). Lastly, a regular hydrogel interface that only comprises stitch linkage corresponds to $h_{\mathrm{pen}}/h_{\mathrm{dc}}\rightarrow \infty$ since $h_{\mathrm{dc}}\rightarrow 0$. Fig \ref{fig2}c shows $h_{\mathrm{pen}}/h_{\mathrm{dc}} \approx 0.2$ when $C = 0.024\%$ and increases to unity as $C$ increases to $0.06\%$ for the TEA gel interface. By tuning $C$, we can vary the degree of engagement and consequently the formation of different linkages, which will be shown later to modulate the resulting adhesion energy.


\section*{Interfaical topological linkages to program rate-dependent adhesion energy}  
To test our hypothesis, we first focus on two extremities: the interfaces containing either slip or stitch linkages. 
To form slip linkage-mediated adhesion, we adhere two TEA gels using chitosan as the bridging polymer with $h_{\mathrm{pen}}/h_{\mathrm{dc}}\approx 0.2$ ($C=0.024\%$), followed by a T-peeling specimen to measure the adhesion energy $G$ as a function of crack speed $V_{\mathrm{crack}}$ (Methods and Fig \ref{figS2}a). 
Fig \ref{fig2}d shows that the slip linkage-mediated adhesion $G^{1/2}$ varies logarithmically with $V_{\mathrm{crack}}$, the crack speed. We observed a factor of 25 in the change of $G$ as $V_{\mathrm{crack}}$ varies by two decades. Together plotted in Fig \ref{fig2}d is the stitch linkage-mediated adhesion formed between two regular hydrogels for the same $C$ and chitosan concentration $c_{\mathrm{chi}}$, showing higher magnitude but much weaker rate-dependence. The contrast between slip and stitch adhesion is the most pronounced at low $V_{\mathrm{crack}}$ but diminishes at high $V_{\mathrm{crack}}$. We also observed adhesive failure and mixed adhesive-cohesive failure at the slip and stitch linkage-mediated interfaces, repspectively.
Our experiments further confirmed the similar bulk mechanics between the TEA and the regular gels: they both show minimal hysteresis in cyclic loadings and weak rate dependences, indicating near-perfect elasticity (Fig \ref{figS1}(a)-(d)). The data suggest that different interfacial network topologies regulate hydrogel adhesion independent of the bulk properties. 

These results motivate us to further analyze the data with a kinetic model proposed by Chaudhary\cite{chaudhury1999rate}. The model considers the breaking of linkages as thermally activated processes\cite{evans1997dynamic,chaudhury1999rate,bell1978models,pobelov2017dynamic}, 
and treats each linkage as a linear spring with stiffness $k_{\mathrm{i}}$ and an activation energy of dissociation $E_{\mathrm{i}}$ (i can be slip or stitch). These parameters influence the dissociation rates of the linkages (Fig \ref{fig2}e), and consequently the rate-dependence of the hydrogel adhesion energy. As detailed in Supplementary note \ref{sec:note2}, the model states the adhesion energy for linkage $i$ relates to the crack speed via $G^{1/2}\sim \ln V_{\mathrm{crack}}$, which agrees perfectly with our experimental data for the slip linkage-mediated adhesion (blue dash lines in Fig \ref{fig2}d).
Further, the model shows that the slope of the linear relation scales inversely to $k_i^{1/2}$, while the intercept depends on $E_i$. By fitting this model to our data we were able to determine $k_i$ and $E_i$, which are otherwise difficult to characterize directly. Specifically, we found  $k_{\mathrm{slip}} =1.7\times 10^{-7}$ N/m and $E_{\mathrm{slip}}$ = 75 kJ/mole for the slip linkage with $C=0.024\%$ and $h_{\mathrm{pen}}/h_{\mathrm{dc}}\approx 0.2$. It is plausible that the hydrogel dangling chains determine $k_{\mathrm{slip}}$ which is of entropic type, so $k_{\mathrm{slip}}\sim k_B T/R^2$ with $k_B T$ the energy in temperature and $R$ the average end-to-end distance of the dangling chains. We estimated that $R\approx 250$ nm, which is 50 times larger than the mesh size of the underlying network $\xi \approx 5$ nm (Supplementary note \ref{sec:note1}). The fitted value of $E_{\mathrm{slip}}$ is larger than the typical activation energy of hydrogen bond (4-50 kJ/mole), suggesting potential synergistic contributions of multiple hydrogen bonds (between chitosan and PAAm) to a single slip linkage. 
Besides, the model captures the rate-insensitivity of $G_{\mathrm{stitch}}^{1/2}$ of the stitch linkage with $k_{\mathrm{stitch}} \geq \sim 300 k_{\mathrm{slip}}$ and $E_{\mathrm{stitch}} \approx 185$ kJ/mol (red dash line, Fig \ref{fig2}b). The much larger $k_{\mathrm{stitch}}$ may be due to the full extension of the entangled networks prior to network rupture, driving the polymer chains far beyond the entropic limit. The estimation of $E_{\mathrm{stitch}}$ is in the range of the bond energy of the C-C bond (350 kJ/mole) \cite{creton2016fracture} and the theoretically estimtated energy stored in each bond prior to rupture using molecular parameters (60 kJ/mole)\cite{wang2019quantitative}, in line with the assumption that the stitched networks must rupture during separation. This model reveals quantitatively that the slip linkages exhibit much lower stiffness and dissociation energy compared to those of stitch linkages.

 Additionally, the model predicts that the hybrid linkage formed when $h_{\mathrm{pen}}/h_{\mathrm{dc}}$ is close to 1, would impart tunable dependence on loading rate through the relation $G_{\mathrm{hybrid}} = G_{\mathrm{slip}}+G_{\mathrm{stitch}}$. In this case, $G_{\mathrm{hybrid}}^{1/2}$ is predicted to be a nonlinear function of $\ln V_{\mathrm{crack}}$ with a finite and constant value of $G_{\mathrm{stitch}}$ (Fig \ref{fig1}c), indicating that the hybrid linkage behaves as slip or stitch linkage respectively in different ranges of loading rates. 
 To test the hypothesis, we prepared TEA gels with $h_{\mathrm{pen}}/h_{\mathrm{dc}}\approx 0.6$ ($C=0.048\%$, Fig \ref{fig2}c), and the resulting $G^{1/2}$ shows a nonlinear trend as expected: at high crack speed, the data collapses onto a master curve with that with $h_{\mathrm{pen}}/h_{\mathrm{dc}}\approx 0.2$ ($C=0.024\%$), following $G_{\mathrm{slip}}^{1/2}\sim \ln V_{\mathrm{crack}}$ (Fig \ref{fig2}f). Note that in this regime, the slip linkage-mediated adhesion is higher than that mediated by stitch linkage for the same $C$ between two regular gels (Fig \ref{fig2}f inset). Below $V_{\mathrm{crack}}$=0.5mm/s, the data converges to a plateau corresponding to rate-independent adhesion energy of $\sim 50$ Jm$^{-2}$. This baseline adhesion is also close to the value of $G_{\mathrm{stitch}}$ for the same $C$ (~60 Jm$^{-2}$, Fig \ref{fig2}f inset), confirming the co-existence of stitch- and slip-linkages on the interface. Fixing $G_{\mathrm{stitch}}$ = 50 Jm$^{-2}$, our model captures the experimentally measured $G_{\mathrm{hybrid}}^{1/2}$ with fitting parameters $k_{\mathrm{slip}} =1\times 10^{-7}$ N/m and $E_{\mathrm{slip}} =71$ kJ/mole (Fig \ref{fig2}f, cyan dot line), closed to the values of the sample with $h_{\mathrm{pen}}/h_{\mathrm{dc}}\approx 0.2$ ($C=0.024\%$). The ability to control the formation of linkage by tuning the entanglement length between TEA gel and bridging polymers offers a high level of adhesion programmability: not only can we predictably tune the adhesion energy by varying loading rates, but also program rate dependence in different ranges of loading rate. The finite adhesion energy at low loading rates provided by the hybrid linkage can effectively prevent the adhesive from failing at static load to ensure good durability. 

 \section*{Programming adhesion kinetics}
In addition to the equilibrium state of adhesion, we next demonstrate that the association of the topological linkages regulates the transient adhesion, which can be exploited to encode adhesion kinetics (Fig \ref{fig1}d). When the bridging polymer solution is placed between the hydrogel and a permeable substrate, they diffuse into the two networks while crosslinking into a bridging network in response to a trigger. The reaction-diffusion process comprises two concurrent sub-processes: the gelation and the diffusion of the bridging polymer with their respective kinetic time $t_{\mathrm{gel}}$ and $t_{\mathrm{d}}$, respectively. We assume that the overall adhesion kinetics is governed by the slower sub-kinetics: $t \equiv \max \{t_{\mathrm{d}}, t_{\mathrm{gel}}\}$. 

When using chitosan as the bridging polymer, the gelation process is due to the pH change in the solution, which is associated with the diffusion of gelling trigger (protons) away from the cast adhesive solution. The thickness of the solution $h_{\mathrm{sol}}$ sets the critical diffusion length, and thus its kinetics time follows $t_{\mathrm{gel}}\sim h_{\mathrm{sol}}^2 /D_{\mathrm{eff,gel}}$\cite{steck2020topological} where $D_{\mathrm{eff,gel}}$ is the effective diffusion coefficient of the gelling trigger. However, $h_{\mathrm{sol}}$ is sensitive to the applied compression or wettability of the interface, yielding the gelation kinetics uncertain in practice without carefully controlled $h_{\mathrm{sol}}$.

\begin{figure}[h!] 
    \centering
    \includegraphics[width=0.7\textwidth]{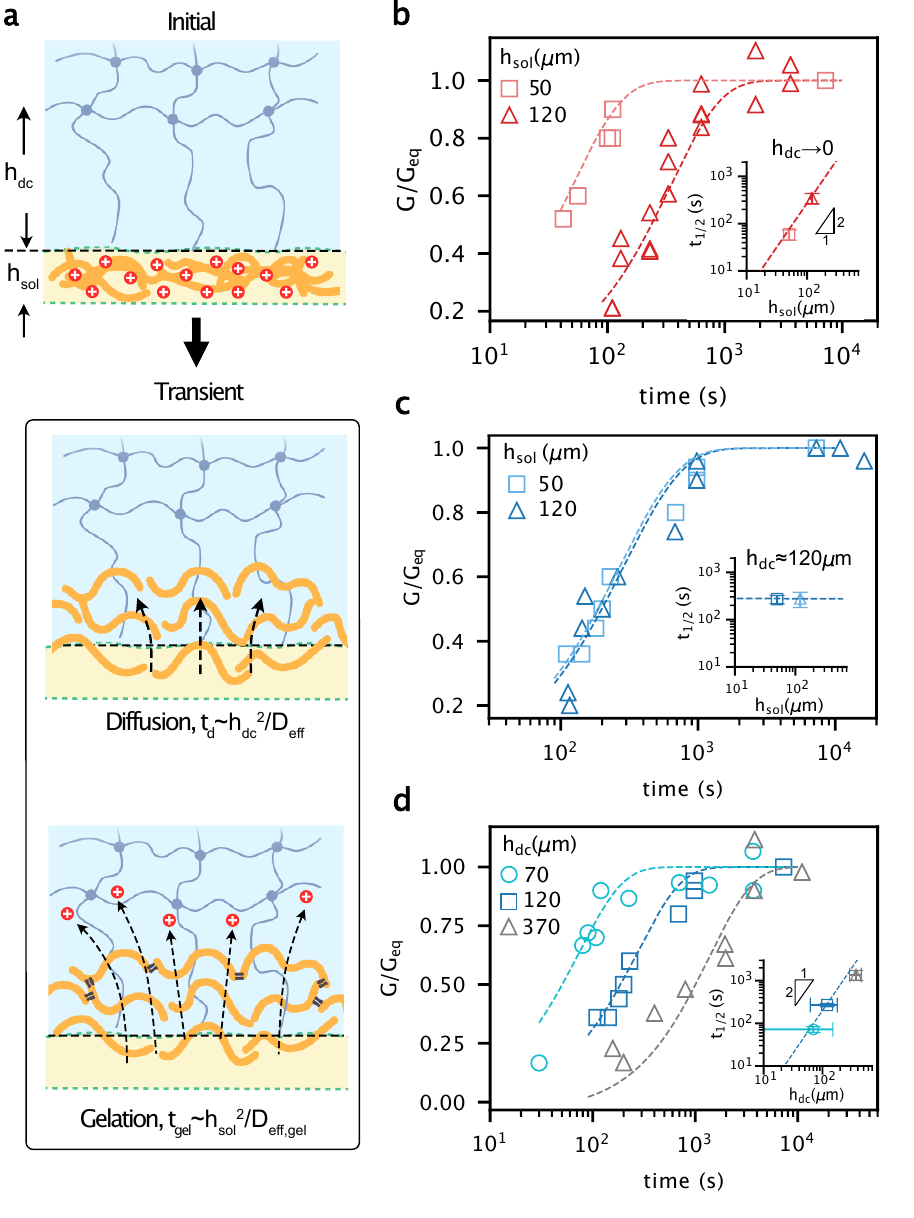}
    \caption{
        \textbf{Programmable adhesion kinetics of TEA.}(a) Illustrations showing that the total adhesion kinetics comprises two sub-kinetic processes: diffusion and gelation. (b) Dimensionless adhesion between two regular hydrogels $G/G_{\mathrm{eq}}$ as a function of waiting time for different cast solution thicknesses $h_{\mathrm{sol}}$. The inset shows $t_{1/2}$ as a function of $h_{\mathrm{sol}}$. Error bars represent 95\% confidence intervals from fitting the exponential function. (c) Similar curves as (b) measured at the interface between two TEA gels with $h_{\mathrm{dc}}\approx 120 \mu m$. (d) Adhesion kinetics of TEA interfaces  with fixed $h_{\mathrm{sol}} $ ($50 \mu m$) and varying values of $h_{\mathrm{dc}}$ ($ h_{\mathrm{dc}}\approx 370, 120, 70 \mu m$, achieved using $C=0.024\%$, $0.048\%$, and $0.06\%$, respectively). The inset shows $t_{1/2}$ as a function of $h_{\mathrm{dc}}$. The y error bars represent a 95\% confidence interval from fitting an exponential function while the x error bars represent the standard deviation from 3 measures.  
    }
    \label{fig4} 
    \end{figure}
    
Meanwhile, the diffusion process of bridging polymers depends on the value of $h_{\mathrm{dc}}$, and hence the type of formed linkages. For a regular gel, $h_{\mathrm{dc}}\rightarrow 0$, the interface is dominated by stitch linkages which only require the bridging polymer to diffuse by one mesh size of the gel network, thus taking negligible kinetic time $t_{\mathrm{d}}\approx 0$ s \cite{steck2020topological}. Thus, one can expect the adhesion kinetics of the regular hydrogel interface to be limited by $t_{\mathrm{gel}}$, which is difficult to control in practice due to the variable $h_{\mathrm{sol}}$. We hypothesize that incorporation of slip or hybrid linkages with dangling chain layers of finite values of $h_{\mathrm{dc}}$ can resolve the issue. In this case, the formation of the linkages requires the bridging polymers to diffuse through the dangling chains layer (Fig \ref{fig4}a). $h_{\mathrm{dc}}$ thus sets the characteristic diffusion length such that $t_{\mathrm{d}}\sim h_{\mathrm{dc}}^2/D_{\mathrm{eff}}$. The prolonged diffusion process can bypass the uncertain gelation process to govern the overall adhesion kinetics. Importantly, since $h_{\mathrm{dc}}$ is a material property, it can render the overall adhesion kinetics insensitive to processing or environmental conditions. 

 
To test the hypothesis, we characterized the adhesion kinetics with different values of $h_{\mathrm{sol}}$ (50 and 120 $\mu$m) controlled by nylon meshes of different thicknesses\cite{steck2020topological} (Methods and  Fig \ref{figS3}). We define the adhesion kinetics using the half time $t_{1/2}$ when $G/G_{\mathrm{eq}}$ reaches 1/2, where $G_{\mathrm{eq}}$ is the adhesion energy in equilibrium. For the regular gel interface, we observe a strong $h_{\mathrm{sol}}$-dependent adhesion kinetics, and the associated kinetic time follows $t_{1/2} \sim h_{\mathrm{sol}}^2$ (Fig \ref{fig4}b and inset). On the contrary, we observe that the adhesion kinetics of the TEA gel interface with $h_{\mathrm{dc}}\approx 120 \mu m$ ($C=0.048\%$) is insensitive to the value of $h_{\mathrm{sol}}$ (Fig \ref{fig4}c and inset). Our point is further strengthed by applying an initial compression (15\% strain) to the TEA gel interface ($h_{\mathrm{dc}} \approx 370 \mu m$) without controlling $h_{\mathrm{sol}}$, which yields the same adhesion kinetics as the TEA gel interface with controlled $h_{\mathrm{sol}}$ (Fig \ref{figS3}c). Thus, incorporation of the engineered dangling chain layer leads to adhesion kinetics insensitive to processing conditions, validating our hypothesis.  

Importantly, not only is the TEA kinetics insensitive to processing conditions but also controllable through changing $h_{\mathrm{dc}}$. Fixing $h_{\mathrm{sol}}$, we observed strong $h_{\mathrm{dc}}$-dependent adhesion kinetics of the TEA gel interface: the kinetics accelerates as $h_{\mathrm{dc}}$ decreases, suggesting shorter distance that the bridging polymers need to diffuse across to form hybrid or slip linkages (Fig \ref{fig4}d). The half time follows $t_{1/2} \sim h_{\mathrm{dc}}^2$ at $h_{\mathrm{dc}}\approx$ 70 and 120 $\mu$m but deviates from the scaling relation at $h_{\mathrm{dc}}\approx$ 370 $\mu$m (Fig \ref{fig4}d inset). In the last case, the kinetics time is presumably bounded by the total diffusion-reaction time since $h_{\mathrm{pen}}/h_{\mathrm{dc}}\ll 1$, indicating that the underlying crosslinked network of the TEA gel is beyond the reach of bridging polymers. The pre-programmable TEA kinetics can be tailored to suit different applications. For instance, a small $h_{\mathrm{dc}}$ can be used with compression to achieve fast kinetics for hemostatic applications\cite{bao2022liquid}, while a large $h_{\mathrm{dc}}$ provides a sufficient time window for adhesive placement.

\section*{Universal applicability}
\begin{figure*}[h!] 
    \includegraphics[width=\textwidth]{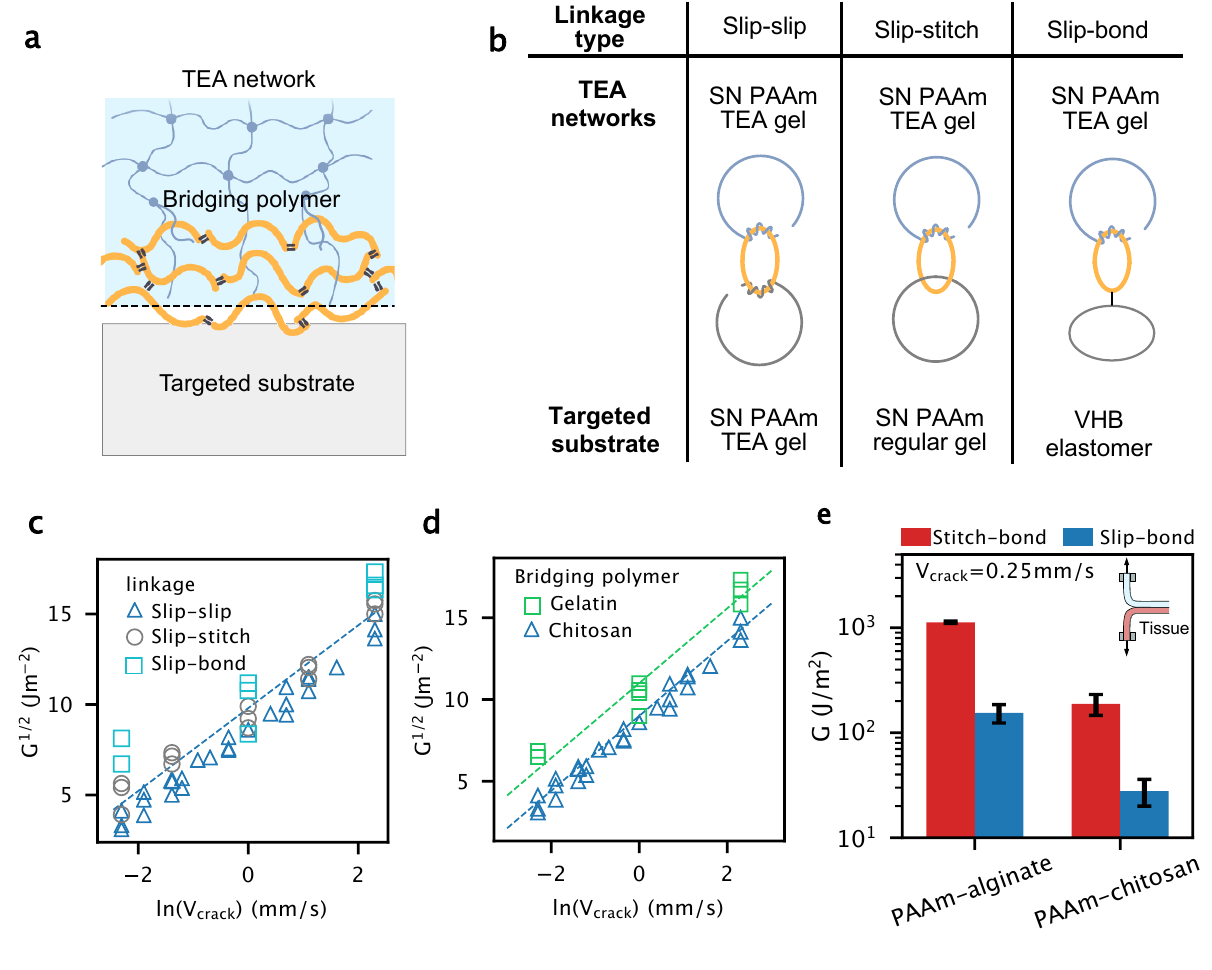}
    \caption{
    \textbf{Universal applicability of the TEA strategy.} 
    (a) Schematic showing that the main constitutes of a TEA interface can be made of a variety of materials. (b) When the adhesive network is made of SN PAAm TEA gel, interfacial topological linkages can be engineered to interact with different targeted substrates. The slip-slip, slip-stitch, and slip-bond linkages are created on targeted substrates TEA gel, regular gel, and VHB elastomer, respectively. (c) Different topological linkages lead to a similar trend of $G^{1/2}$ as functions $V_{\mathrm{crack}}$. The SN TEA gel matrix and the chitosan solution are prepared with $C=0.024\%$ and $c_{\mathrm{chi}} = 2\%$ g/ml. (d) Slip-slip mediated adhesion between two TEA gels using gelatin and chitosan as bridging polymers. $c_{\mathrm{chi}} = c_{\mathrm{gelatin}} = 2\%$ g/ml (d) Adhesion of TEA and regular DN gels on porcine skin at relatively low $V_{\mathrm{crack}}$. 
    }
    \label{fig3}
    \end{figure*}

The design and fabrication of TEA are universally applicable to a wide range of material systems (different bridging polymers, targeted substrates, and TEA networks) (Fig \ref{fig3}a). Slip linkages formed on the gel-bridging network interface could be coupled with other interactions such as slip, stitch linkages or covalent bonds \cite{Yang2019} that the bridging network can interact with the targeted substrates. For instance, the triggered crosslinking and the abundant amino groups of chitosan provide numerous options to interact with different substrates through covalent or physical interactions\cite{yang2019design}. Based on the principle, slip-slip, slip-stitch, and slip-bond linkages were achieved between two TEA gels, between a TEA and a regular gel, and between a TEA gel and a VHB elastomer, respectively (Fig \ref{fig3}b, Table \ref{tab1}). Remarkably, our data reasonably collapse for the different mechanisms to engage different targeted substrates (Fig \ref{fig3}c), suggesting that the overall adhesion behavior is dictated by the slip linkages while depending less on the types of interactions between the bridging network and targeted substrates. These results support the robustness of the adhesion programming through the TEA strategy.

Besides chitosan, we examine another bridging polymer gelatin, which was prepared as polymer solution at $37^{\circ}C$ and then applied to the interface between two TEA gels for $C=0.024\%$ at room temperature. Similar to chitosan, gelatin diffused into the gel and was crosslinked into a bridging network in responding to a temperature drop to form slip linkages with the TEA dangling chains. Our data reveals an identical trend between the data obtained using gelatin and chitosan as bridging polymers (Fig \ref{fig3}d), highlighting the dominating role of polymer topology rather than material chemistry in the formation of slip linkages.

We then explore using double-network (DN) hydrogel as the TEA network. Compared to single-network (SN) hydrogels, the DN hydrogels exhibit much higher fracture toughness and adhesion\cite{Yuk2016,Sun2012,Li2017,bao2021ionotronic} due to background dissipation. We tested PAAm-alginate and PAAm-chitosan hydrogels as representative materials. In the two types of DN gels, alginate and chitosan are physically crosslinked macromolecules and do not covalently interfere with the PAAm network, we expect that the hydrophobic mold could produce surface dangling chains in the PAAm network within the DN gels. We confirmed the presence of the dangling chain layer in the surface of a PAAm-alginate hydrogel polymerized on hydrophobic substrate by EDTA treatment to remove calcium-alginate bonds followed by Atomic Force Microscopy (AFM) tests (Fig \ref{figS5}a, b). 
We then examined the adhesion of TEA and regular DN gels polymerized on PMMA and glass molds on porcine skin (for a systematic study on different gelling molds, see Fig \ref{figS5}c). We use chitosan as the bridging polymer and EDC/NHS reagent to form covalent bonds between chitosan and tissue surfaces \cite{Li2017}. 
Since the energy dissipation in the DN gels is coupled to the interfacial adhesion energy, we hypothesized that the slip linkage elicits negligible bulk energy dissipation at low $V_{\mathrm{crack}}$, thereby resulting in weak adhesion. Here, we only focused on the adhesion behavior of the DN TEA at low $V_{\mathrm{crack}}$, since its rate-sensitive adhesion is presumably coupled with the rate-dependent bulk dissipation and is not further pursued \cite{Yang2018}. 
Both PAAm-alginate and PAAm-chitosan gels show slip linkage-mediated adhesion 10 times lower than stitch-mediated adhesion at $V_{\mathrm{crack}}$=0.25mm/s (Fig \ref{fig3}e), demonstrating that our methodology is applicable to both SN and DN hydrogels as long as the topology of one of the networks can be engineered. 

\section*{Programming spatial adhesion}
\begin{figure}[H] 
    \centering
    \includegraphics[width=0.95\textwidth]{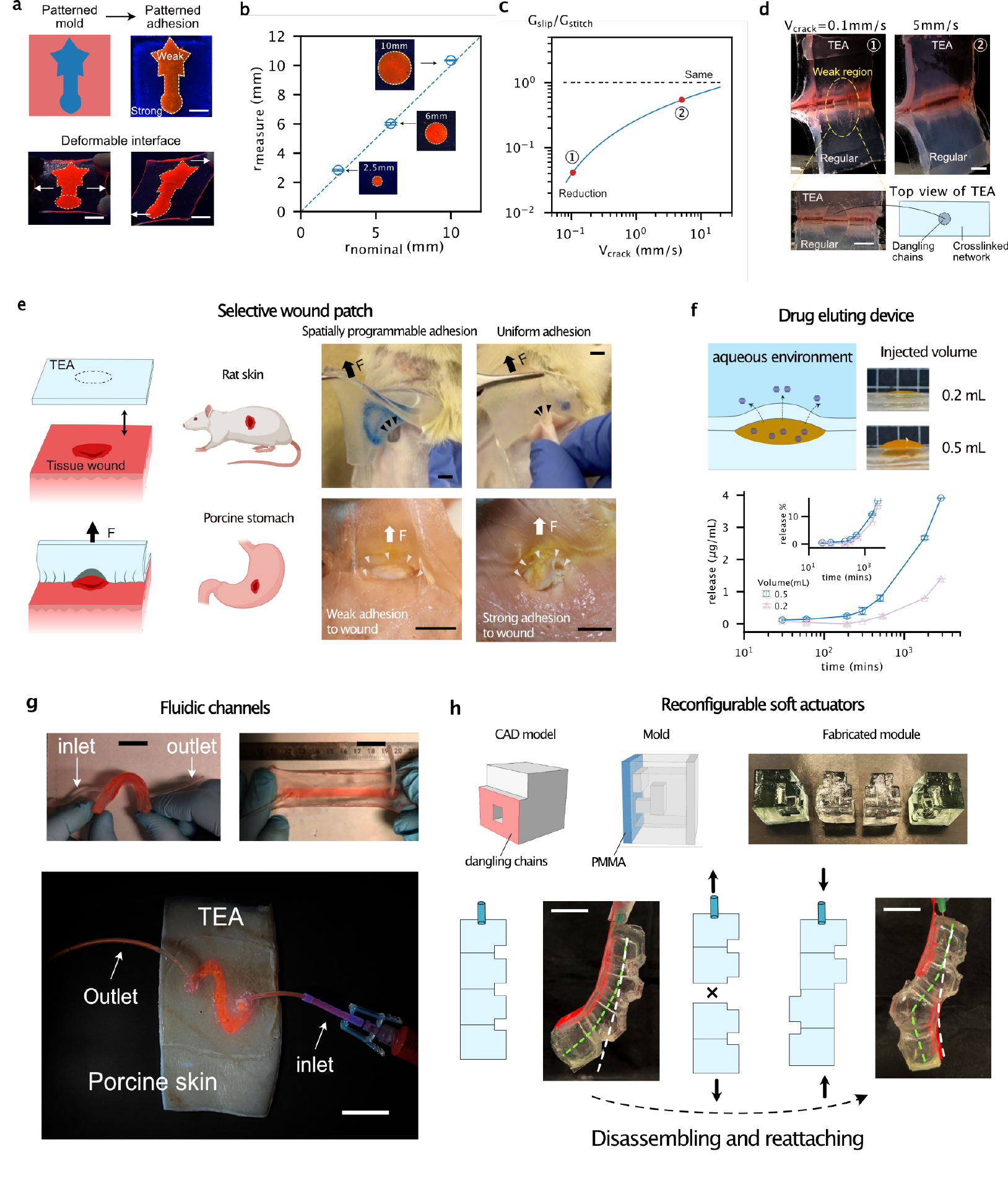}\centering
    \caption{
        \textbf {Spatial programming and soft devices enabled with TEA}
        (a) TEA strategy leads to spatially programmable and deformable adhesion capable of tracing complex shapes between a PAAm-alginate TEA DN gel and a regular DN gel. Scale bar: 1cm. (b) Spatial resolution of the spatially programmable adhesion. (c) Predicated $G_{\mathrm{slip}}/G_{\mathrm{stitch}}$ achieved using the parameterized model (Eqn.\ref{eqn:G_final}) using the data of $G_{\mathrm{slip}}$ and $G_{\mathrm{stitch}}$ in Fig \ref{fig2}d. (d) Experimental demonstration of the rate-dependent $G_{\mathrm{slip}}/G_{\mathrm{stitch}}$ between a TEA gel with a circular-shaped dangling chain region and a regular gel. Scale bar: 1cm. (e) Wound patches made of TEA and regular PAAm-alginate DN gel adhered to wounds on rat skin (top, scale bar: 8mm) and porcine stomach (bot, scale bar: 12mm). (f) A drug-eluding device enabled by injecting drug into the weakly-adhered interface between a TEA gel and a regular gel. Grid size of the inset: 10mm. (g) Deformable hydrogel-based fluidic channels created by adhering a PAAm-alginate TEA DN gel to a regular DN gel. Scale bars: 2cm. (Bottom) A PAAm-alginate TEA DN gel with designed adhesion selectivity forms a fluid channel on the surface of porcine skin. Scale bar: 2.5cm. (h) Reconfigurable soft actuators. (Top) the fabrication process of the actuator units with connection surfaces composed of dangling chains. (Bottom) two modes of actuation. The initial and the actuated stages are indicated by the white and green dash lines, respectively. Scale bars: 2cm. 
    }
    \label{fig5}
    \end{figure}

The contrast between slip and stitch linkages allows us to program the adhesion spatially. To do so, we patterned a mold substrate with hydrophilic (Glass) and hydrophobic regions (PTFE films thickness $\sim$0.1mm), followed by polymerizing a TEA gel on the patterned mold. Given the predefined geometries (circle, triangle) of the hydrophobic domains, we can design the dangling chain region where weak adhesion $G_{\mathrm{slip}}$ is formed at low loading rates; meanwhile, strong adhesion $G_{\mathrm{stitch}}$ is formed in other areas to sustain tension or twisting applied to the interface without interface debonding. Fig \ref{fig5}a and \ref{figS5}e show that by shaping the dangling chain region, we can achieve weak adhesion region of complex shapes between a TEA gel and a regular gel by slowly injecting liquid dye into the weak interface. 
To further characterize the resolution of the spatially programmable adhesion, we made a series of circular islands of nominal radii $r_{\mathrm{nominal}}$ in which slip linkages are formed. By slowly injecting the liquid dye, we visualized and measured their radii $r_{\mathrm{measure}}$ of the weak adhesion region using a digital camera (Fig. \ref{fig5}b and \ref{figS5}c). The excellent agreement between the nominal and measured radii suggests high spatial resolution $\sim$0.1 mm achieved with a manual procedure.

As the slip and stitch linkages show different sensitivities to loading rate, we expect rate-dependent spatially programmable adhesion, characterized by the adhesion energy contrast $G_{\mathrm{slip}}/G_{\mathrm{stitch}}$. Fig \ref{fig5}c shows that $G_{\mathrm{slip}}/G_{\mathrm{stitch}}$ predicted by the parameterized model (Supplementary note \ref{sec:note2}) approaches unity at high $V_{\mathrm{crack}}$ and decreases towards zero at low $V_{\mathrm{crack}}$.  The prediction is supported by our experimental observations: A TEA gel with the designed dangling chain region shows large adhesion contrast to a regular gel at low $V_{\mathrm{crack}}$ while appearing to be uniformly adhesive at relatively larger $V_{\mathrm{crack}}$ (Fig. \ref{fig5}d). The rate-dependent spatially-programmable adhesion can potentially enable applications which desire tunable adhesion contrast in different regions under different loading rates. Additionally, not only we can achieve reduced adhesion ($G_{\mathrm{slip}}/G_{\mathrm{stitch}}<1$) but also enhanced adhesion ($G_{\mathrm{hybrid}}/G_{\mathrm{stitch}}>1$) in the region with slip linkages by leveraging the hybrid linkage at large $V_{\mathrm{crack}}$ (Fig \ref{figS4}). In this case, the slip linkage acts as a toughening mechanism that synergistically contributes to the adhesion unit with the stitch linkage. Moreover, the one-step fabrication allows spatially selective adhesion to be assembled within a monolithic material, which otherwise requires assembling different materials at the interface. 

\newpage

\section*{TEA-based devices}
The programmable adhesion of TEA enables various applications such as wound patches, drug depots, fluidic channels, and soft actuators. For the application of wound patches, TEA allows one to program weak adhesion to wound beds while maintaining strong adhesion to the surrounding tissue. As such, the patch could protect the wound without impairing tissue regeneration and wound closure. Using the one-step fabrication process (Fig \ref{fig1}e and \ref{figS5}d), we prepared such a TEA gel with its surface composed of a circular region of dangling chains and the surrounding region of crosslinked network.  The dangling chain region forms slip linkages which attach weakly to the wound site upon slow removal to minimize the damage to the wound. Meanwhile, the stitch linkages attach strongly to the surrounding healthy tissue to maintain the stickiness of the patch. In contrast, a regular hydrogel exerts strong and uniform adhesion to both wounded and healthy tissues, which ruptures the wound bed upon removal (Fig \ref{fig5}e). 

Besides, the creation of a weak adhesion region between two hydrogels could serve as a drug depot. Upon slow injection, mock drug solution filled up the weak interface. Further injection created a bulge of hydrogel to accommodate a high amount of drug, which can be continuously released through the hydrogel network when the whole device is immersed in an aqueous environment (Fig. \ref{fig5}f Top). Our data show that the initial amount of drug injected into the depot affects the amount of release over time but the relative kinetics of release remains similar (Fig. \ref{fig5}f Bottom). As well, we can create a drug depot above a wound site, where drugs can be directly released into wounded tissue. In contrast, the strong adhesion of a regular hydrogel prevents the injection of drug solution to interface (Fig. \ref{figS5}f).

We then demonstrate hydrogel-based fluidic devices assembled with TEA. A PAAm-alginate TEA DN gel with a rectangular-shaped dangling chain region forms a partially weak interface with a regular DN gel, which subsequently becomes a fluidic channel upon slow injection of liquids. The resulting device is highly deformable while no liquid leakage is observed (Fig \ref{fig5}g Top). The one-step fabrication technique provides a simple approach to fabricate hydrogel fluidic channels compared with conventional methods that typically involve multiple molding steps \cite{gjorevski2022tissue,lin2016stretchable}. In addition, the spatially programmable adhesion is applicable to varying surfaces as it requires no patterning of the targeted substrate. As such, we can form such a fluid channel directly on tissue surfaces such as porcine skin (Fig \ref{fig5}g Bottom). This feature could benefit medical devices that contact with tissue surfaces for sustained drug release \cite{whyte2018sustained}, or in-\textit{vitro} organ-on-chip models to study cellular behaviors \cite{vera2021engineering}.

Lastly, we show that the TEA made of SN PAAm hydrogels can be used to construct reconfigurable soft actuators, featuring minimal bulk dissipation for efficient actuation and dynamic adhesion for reversible attachment (Fig \ref{fig5}h). Such actuators are formed with hydrogel units that contain surface dangling chains on each face and are connected to each other with the aid of bridging polymer (Methods). The slip linkage-mediated adhesion between the units is strong enough to sustain actuation, and yet can be separated easily and slowly with a small force. The separated units can then be reconnected upon reapplying the bridging polymer solution to the interface so that one can modify configurations of assembly for different actuation. Our data shows that the slip linkage-mediated adhesion increases and reaches a plateau after cycles of detachment and reattachment (Fig \ref{figS2}f). This property can be partially attributed to the fact that the dissociation of the slip linkage does not rupture the adherend networks, so the slip-mediated TEA interface is inherently subjected to minimal damage compared with those bonded by stitch linkages or covalent bonds (Fig \ref{figS2}g).

\section*{Conclusion}
In summary, we have demonstrated that designing the interfacial network topologies of hydrogels provides a facile and robust approach to program adhesion in multiple aspects including magnitude, space, and kinetics. 
Our approach can be potentially extended to different length scales using proper manufacturing processes. For instance, spatially programmable adhesion with a spatial resolution on the micro-scale can be achieved with microfabrication of the hydrogel network topology \cite{tudor2018fabrication,carlotti2019functional}, while that on the metre scale is expected to be achieved using gelling molds of the same size for applications such as camouflaging skin \cite{pikul2017stretchable}. 
Broadly, our methodology falls into the emerging paradigm of material intelligence, as the adhesion programming is directly encoded in the hydrogel network as material properties, similar to other properties such as elastic modulus. The implementation of adhesion control requires no external apparatus, making the methodology extremely facile, robust, and scalable.
We hope that the design of TEA can spark interest in controlling hydrogel adhesion by designing their network topologies, opening the door to a new design space for intelligent materials/structures through programmable adhesion.

\bibliography{library}

\section*{Acknowledgments}
This work is supported by the Natural Sciences and Engineering Research Council of Canada (grant RGPIN-2018-04146), Fonds de Recherche du Quebec-Nature et technologies (grants FRQNT PR-281851), and the National Institute on Deafness and Other Communication Disorders (grant R01-DC018577). Z.Y.  and R.H. acknowledge support from FRQNT and McGill Engineering Doctoral Award. S.J. acknowledges support from FRQNT Doctoral Award. X.Y. and R.L. acknowledge support from the National Science Foundation of the United States (NSF CMMI-1752449). J.L. acknowledges the support from the Canada Research Chair Program. The authors acknowledge Mr. Tianqin Ning and Mr. Yixun Chen for taking pictures, Mr. Omar Peza-Chavez for the help with cryosectioning, and Prof. Reghan Hill for valuable discussion.

\section*{Competing financial interests}
The authors declare no competing financial interests.

\section*{Methods}\label{Methods}
\textbf{Materials} All chemicals were purchased and used without further purification. Materials for the hydrogel synthesis and the bridging polymer include acrylamide (AAm, Sigma-Aldrich, A9099), N,N’-methylenebisacrylamide (MBAA; Sigma-Aldrich, M7279), ammonium persulphate (APS, Sigma-Aldrich, A3678), N,N,N’,N’-tetramethylethylenediamine (TEMED, Sigma-Aldrich, T7024), Alginate (I-1G) was purchased from KIMICA Corporation, chitosan (degree of deacetylation, DDA: 95\%, medium and high molecular weight, Lyphar Biotech), sodium bicarbonate (Fisher Scientific, S233), sodium phosphate monobasic (NaH2PO4, Sigma, S8282), sodium phosphate dibasic (Na2HPO4, Sigma-Aldrich, S7907), acetic acid (Sigma-Aldrich, A6283), calcium sulfate (Sigma-Aldrich), N-hydroxysulfosuccinimide     (NHS, Sigma-Aldrich, 130672),     and     1-ethyl-3-(3-dimethylaminopropyl) carbodiimide (EDC, Sigma-Aldrich, 03450), Gelatin (Sigma-Aldrich, G2500). Glass, acrylic sheets (PMMA), PS, silicon, and PTFE were purchased from McMaster-Carr to make mold substrates for polymerization.  VHB elastomer was purchased from 3M. Porcine skin was purchased from a local grocery store, then stored in a fridge at -20°C, and thawed at 4°C before use. Nylon mesh were purchased from McMaster Carr without further modification (9318T25, 9318T23 for thicknesses of 50 $\mu m$ and 120 $\mu m$, respectively.) 

\textbf{Synthesis of TEA} The TEA based on PAAm hydrogels was prepared as follows. AAm monomers of 6.76 g was first dissolved in 50 mL of deionized water. After degassing, the AAm solution of 25 ml was mixed with varying amounts of MBAA aqueous solution (0.02 g mL$^{-1}$) and 20 $\mu L$ of TEMED in a syringe. The volumes of MBAA solution added were 90, 120, 150, 180, and 240 $\mu L$ for the crosslinker$-$to$-$monomer molar ratios $C$ at 0.024$\%$, 0.032$\%$, 0.04$\%$, 0.048$\%$, 0.06$\%$, respectively. Meanwhile, another syringe was added with 565 $\mu L$ of APS solution (0.066 g mL$^{-1}$) and 478 $\mu L$ deionized  water. The two syringes were connected with a Luer$-$Lock connector, so the two solutions were syringe$-$mixed to form a homogeneous solution. The mixture was immediately injected into rectangular acrylic molds of 80 $\times$ 20 $\times$ 3 mm$^3$ or 80 $\times$ 15 $\times$ 1.5 mm$^3$, covered with PMMA on both sides, and then kept at room temperature for 24 hours to complete the reaction. To prepare the regular PAAm gels, we follow the same procedure except for injecting the mixed solution into the acrylic molds covered by glass sheets on the two sides.

The TEA based on double-network gels was prepared with the following protocol. To prepare PAAm-alginate TEA, 1.5g alginate (I-1g) power and 6.76 g AAM were dissolved in 50 mL of deionized water. The first syringe was prepared following the aforementioned protocol, while the second syringe was added with 565 $\mu$L of APS solution and 478 $\mu$L calcium sulfate solution (15\% w/v). The precursor solutions were quickly syringe-mixed and immediately poured into 80 $\times$ 20 $\times$ 3 mm$^3$ rectangular acrylic molds covered with PMMA on the two sides, and then kept at room temperature for 24 hours to complete the reaction. To prepare the PAAm-Chitosan TEA, acrylamide and chitosan powders were first dissolved in 0.2 M acetic acid at 3.3 mol/L and 2.5\%, respectively. MBAA was then added to the AAm-chitosan solution at 0.0006:1 the weight of acrylamide to complete the polymer precursor solution. To prepare a gelling solution to crosslink the chitosan, 0.1 M Na$_2$HPO$_4$ and 0.1 M NaH$_2$PO$_4$ were first mixed with a volume ratio of 50:3. Sodium bicarbonate was then added to the solution at a concentration of 0.306 M. A mass fraction of 6.6\% APS was later added to the solution as an initiator. Both solutions were degassed, quickly mixed at 3:2 volume ratio (polymer precursor to gelling solution) using syringes and injected into a mold with substrates of choice for overnight gelation. To prepare the regular PAAm-alginate and PAAm chitosan gels, we follow the same procedure except for injecting the mixed solution into the acrylic molds covered by glass sheets on the two sides followed by the gelation process.

\textbf{Preparation of the bridging polymer solutions.} This study tested two types bridging polymers: chitosan and gelatin. To prepare the chitosan solutions of 2\%, 1\%, and 0.5\% w/v, 50 mL of deionized water was added with chitosan powders of 1, 0.5, 0.25 g, respectively. 400 $\mu L$ of acetic acid was also added for a final pH of 4.5. The mixture was stirred overnight to form a homogeneous solution and then kept at 4$^{\circ}$C before use. To prepare the gelatin solution of 2\% w/v, 1g of gelatin powder was dissolved in 50 mL of deionized water. The solution was stirred in a water bath at 37$^{\circ}C$ for 30 mins before use.

\textbf{Adhesion test}. The TEA gels were prepared with a length of 80 mm (or 40 mm), width of 20 mm, and thickness of 3 mm. To test their adhesion on gels and VHB elastomers, the surface of TEA was treated with the bridging polymer solution of 0.25 $\mu$L/mm$^2$, and then immediately covered with the adherend. An instant compression of 15\% strain was applied to remove the excessive solution on the interface. No prolonged compression was applied. For testing TEA adhesion on tissue samples, bridging polymer solution with chemical reagents added was applied to the tissue surface, followed by covering with the TEA gel. A continuous compression of 15\% strain over the whole course of adhesion establishment was applied to the TEA-tissue sample. To measure the adhesion energy, standard 180-degree peeling was performed based on on samples with 2 cm width. The tests were conducted using a universal testing machine (Instron Model 5365) with 10N and 1kN load cells. Before test, PET backing is attached to the samples using Krazy glue. In a typical test, the peeling force reaches a plateau $F_{\mathrm{plateau}}$ once reaching the steady state process. The adhesion energy is calculated as twice the plateau force divided by the sample width, $G = 2F_{\mathrm{plateau}}/w$. The loading rate was varied from 0.2 mm/s to 40 mm/s. Given the rigid backing, the crack speed for 180-degree peeling is half the loading rate.

\textbf{Adhesion kinetics test.} To characterize adhesion kinetics, nylon mesh of different thicknesses (50 and 120 $\mu m$) were used to define the thickness of the bridging polymer solution $h_{\mathrm{sol}}$ following a previously reported protocol\cite{steck2020topological}. We first immersed the nylon mesh in the bridging polymer solution, and then removed the excessive solution on the surface before applying it to the interface between two pieces of hydrogels. We waited for different time $t$ before measuring adhesion energy using T-peeling test at a relative low crack speed $V_{\mathrm{crack}}=$ 0.25 mm/s. The adhesion energy increases with $t$ and reaches the equilibrium value $G_{eq}$. The adhesion energy versus waiting time was fitted by the function $G=G_{eq} (1-e^{t/t_{1/2}})$, where the half time $t_{1/2}$ can be fitted.

\textbf{Uniaxial tensile test.} Samples with length of 40mm, width of 20mm, thickness of 3mm were prepared and tested using the Instron machine with 1N and 1kN load cells. The nominal stress is calculated using $\sigma=F/A$ where $F$ and $A$ are the measured force and cross-sectional area of the sample. The stretch is calculated as $\lambda=\lambda_1/\lambda_0$ where $\lambda_1$ is the current length and $\lambda_0$ is the initial length.

\textbf{Atomic force microscopy.} An atomic force microscope (JPK NanoWizard@3, Berlin, Germany) was used to conduct nano-indentation tests. Rectangular silicon cantilevers with 0.6 $\mu m$-in-diameter spherical beads attached as probes were used (Novascan, IA, USA). Cantilevers with a nominal spring constant of 0.6 N/m were used for experiment. The cantilever spring constants were determined using thermal noise method before the experiment. Hydrogel samples were immersed in PBS for 3 hours before indentation to avoid fast swelling during test. Then swollen hydrogels were glued to 35-mm Petri dishes and immersed in PBS during the measurement. Hertzian contact model was used to fit the indentation data and to calculate the Young's modulus.

\textbf{GPC test.} Chitosan samples with a concentration of 1 mg/mL were dissolved in an aqueous acetate buffer (0.25 M acetic acid, 0.25 M Na acetate) overnight, with stirring. They were filtrated (0.45 $\mu$ m) prior to be transferred to the autosampler vials. 100 $\mu$L injection was analyzed using a SEC column OHpak SB-805 HQ (8 mm ID x 300 mm L) at a  flow  rate of 0.3 mL /min of the eluting phase (same as the one in which samples were dissolved). The signal of the RI detector was used for calculation of the molecular weights and PDI against a calibration curved constructed with pullulan standards of narrow polydispersity and molecular weight in the 5.9 to 788 kDa range.

\textbf{Confocal microscopy.} Hydrogel samples with diffused bridging polymers were imaged with a confocal microscope (LSM 710, Zeiss). Hydrogel samples with diffused FITC-chitosan were cut into thin slices by a blade and transferred to a 35 mm Petri dish with coverslip bottom (MatTek, P35G-0-10-C). The polymer network was imaged with 10X and 20X objective lenses. Axiovert A1 inverted microscope (Zeiss) equipped with a motorized stage was used to obtain fluorescent signals at multiple locations. Light intensity was maintained the same for all the samples. 

\textbf{Patterned TEA.}
A PTFE tape of thickness 0.1mm was designed into different shapes and then adhered to a glass plate. A rectangular acrylic mold was placed on top of patterned substrate. After injecting pre-gel solution, the mold was covered with another glass plate.

\textbf{TEA-based wound patch.}
A PTFE tape of thickness 0.1mm was designed into a circle of diameter 1.5 cm, and then adhered to a glass plate. A rectangular acrylic mold was placed on top of patterned substrate. After injecting pre-gel solution, the mold was covered with another glass plate. Biopsy punch was used to generate circular wounds on rat skin (diameter of 6mm) and porcine stomach (diameter of 1cm). When de-molded, the TEA gel was applied to wounds on rat skin and porcine stomach with 2\% chitosan and EDC/NHS reagents. A prolonged compression for 3 mins was then applied by hand. The size of the circular region can be tailored to suit different sizes of the wounds.

\textbf{TEA-based fluidic channels.}
A PTFE tape of rectangular (or ‘S’-shape) was attached to a glass plate. Pre-gel solution is injected into the mold for reaction to complete. Once de-molded, we use hole punch of diameter 2 mm to create an inlet hole and an outlet hole, and then attach it to a targeted surface using chitosan as bridging polymer. If the targeted surface is tissue, we also added chemical reagents to form covalent bonds between chitosan and tissue surfaces. After the adhesion is established, we further insert two soft tubes into the inlet and outlet, sealed using Crazy glue, and slowly inject liquid to cleave the weak adhesion by slip linkages. The fluidic channel is formed once the whole weak interface is separated.

\textbf{TEA-based Drug eluting devices.}
A SN PAAm TEA gel was first prepared on a glass mold substrate patterned with a circular PTFE film of diameter 2 cm. It was then adhered to a regular gel which was glued onto an acrylic sheet. The circular low-adhesion region served as local depot to which drugs will be injected. The albumin-FITC (A9771, Sigma-Aldrich) was used as a model drug and was dissolved in PBS at 10 mg/mL to get the drug solution. At equilibrium of hydrogels, different volume of drug solution (0.2 mL and 0.5 mL) was slowly injected into the weak interface, resulting in local drug punches inside the hydrogel. After the injection, the whole device was immersed in 200 mL PBS. At determined time points, the fluorescence intensity of the solutions was measured using a BioTek Synergy HTX multi-mode microplate reader ($\lambda$ex = 485/20 nm, $\lambda$em = 528/20 nm). A standard calibration curve was made in order to calculate the concentration of released drug. 

\textbf{TEA-based reconfigurable soft actuators} 
Individual actuator module was prepared using a molding process. SN PAAm pregel solution was injected to a 3D printed mold made of PLA (Fig \ref{figS6}). The mold was then covered by a PMMA sheet. Once the reaction is completed and demolded, the actuator module readily carries dangling chains on its surface. Different modules were connected by applying bridging polymer chitosan to the interface. To separate the connected modules, a small force with a low separate rate can be applied. The separated modules can be reconnected following reapplying bridging polymers to the interface.

\setcounter{section}{0}
\section{Supplementary Note 1: TEA gel prepared by heterogeneous polymerization.}\label{sec:note1}
We consider here the heterogeneous structure of the TEA gel imparted by the hydrophobic mold substrate. The phenomenon was first discovered by Gong et al.\cite{Gong2001,kii2001heterogeneous,gong2001synthesis}, who proposed that the hydrophobic mold substrate suppresses the polymerization of the precursor solution near the substrate-solution interface due to the increased interfacial tension. When the pre-gel solution starts polymerizing on the low-surface-tension mold such as PMMA, the interfacial tension between the hydrophobic mold surface and the liquid solution $\gamma_{sl}$ increases with the increased polymer fraction. To minimize the free energy of the whole system, the mold surface repulses the polymers to create a depletion layer of thickness $\xi_{dp}$, where the polymer fraction is significantly reduced compared to that in the bulk. As the polymerization continues, the polymer chains in the bulk start to entangle and from a crosslinked network with the bulk network elasticity $E_{bulk}$. Within the depletion layer, however, the crosslinking process is influenced due to the extremely low polymer fraction, resulting in a layer of dangling chains of thickness $h_{dc}$. We shall consider $\xi_{dp}$ and $h_{dc}$ to be equivalent. In equilibrium, the increased interfacial energy $\Delta \gamma_{sl}$ equals the work done against $E_{bulk}$ over the distance $\xi_{dp}$, leading to the relation $\Delta \gamma_{sl} \sim E_{bulk}\xi_{dp}$ or $\Delta \gamma_{sl} \sim E_{bulk}h_{dc}$. 
By assuming $\gamma_{sl}$ changes negligibly with $E_{bulk}$ when $C$ is varied, and all the crosslinkers effectively contribute to load-bearing chains, it leads to the scaling relation: $h_{dc}\sim 1/E_{bulk}\sim 1/C$, which is plotted in Fig \ref{fig2}b with the estimated $h_{dc}$ values. 

Recently, several results suggested that the inhibition of free-radical polymerization by trapped oxygen on hydrophobic substrates may play significant roles in the heterogeneous polymerization\cite{mandal2021oxygen,zhang2019creating}. Despite the debates on the underlying mechanisms, the substrate effect effectively produces heterogeneous structure of a TEA gel that comprises a layer of branched dangling chains spanning a thickness of $h_{dc}$ near the surface, and a homogeneously crosslinked bulk network that is affected minimally by the substrate effect of thickness $h-2h_{dc}$.

\subsection*{Estimation of $h_{\mathrm{dc}}$}
We propose a simple model to characterize the value of $h_{\mathrm{dc}}$ of TEA gels. The model considers the TEA gel has an effective modulus of $E_{tea}$, composed of the bulk elastic modulus $E_{bulk}$ and the dangling chain elastic modulus $E_{dc}$. It assumes that the bulk of the TEA gel the same elasticity $E_{bulk}$ as that of a regular gel polymerized on glass $E_{reg}$, while the dangling chains has a modulus $E_{dc}\approx 0$, yielding:

\begin{equation}\label{eqn:hdc}
\frac{2h_{\mathrm{dc}}}{h}=1-\frac{E_{\mathrm{tea}}}{E_{\mathrm{reg}}}
\end{equation}
where $E_{\mathrm{tea}}$ and $E_{\mathrm{reg}}$ can be measured from uniaxial tensile tests. We fit the data with $5\%$ of strain to the linear-elastic model, which is degenerated to from the neo-Hookean model at the small strain limit:
\begin{equation}\label{eqn:2}
\sigma_{\mathrm{linear}} = E(\lambda-1)
\end{equation}
where $\sigma$ is the nominal stress and $\lambda$ is the stretch. We also fit the data to another hypereasltic material model, the incompressible neo-Hookean model:
\begin{equation}\label{eqn:2}
W_{\mathrm{neo-Hookean}} = \frac{E}{6}(I_1-3)
\end{equation}
where $I_1 = \lambda_1^2+\lambda_2^2+\lambda_3^2$. The Neo-Hookean model assumes that the polymer chains follow Gaussian distribution and are free of entanglements. Under uniaxial tensile test, the principle stretches: $\lambda_1 = \lambda$, $\lambda_2=\lambda_3=1/\sqrt{\lambda}$. The nominal stress-stretch curves derived from the models are:
\begin{equation}\label{eqn:2}
\sigma_{\mathrm{neo-Hookean}}=\frac{E}{3}(\lambda-\lambda^{-2})
\end{equation}

We plot representative stress-stretch curves from uniaxial tensile tests in Extended Data Fig 1d and along with the fitted models. The Neo-Hookean model underestimates the slopes of the curves at $\lambda \rightarrow 1^{+}$ for all samples. The linear model that fit the data within $\lambda$ from 1 to 1.05 is used to estimate the moduli of TEA and regular gels with varying values of $C$, plotted in Extended Data Fig \ref{fig1}e. With the values of $E_{\mathrm{tea}}$ and $E_{\mathrm{reg}}$, we estimated the value of $h_{dc}$ using the Eqn. \ref{eqn:hdc} (Fig 2c). Note that our estimation of $h_{\mathrm{dc}}$ is much larger than those estimated by observing force-displacement curves from nano-indentation tests\cite{mandal2021oxygen}\cite{chau2021simple}, which may attribute to the different values of $C$ used in the studies. The projection of our data yields a similar estimation of $h_{\mathrm{dc}}$ at the same level of $C$ as in Simic et al\cite{mandal2021oxygen} ($C\approx 1\%$).

\subsection*{Estimation of bulk mesh size}
Assuming the bulk network of the TEA gel and that of a regular gel have affine structures, so their shear modulus can be expressed as
\begin{equation}\label{eqn:2}
\mu = \frac{E}{3}=\nu k_B T
\end{equation}
where the density of the network strands $\nu=\frac{c_{\mathrm{aam}}}{m_{\mathrm{mono}}N_c}N_A$.
The concentration of the aam solution is $c_{\mathrm{aam}}=6.76\mathrm{g}/50\times 10^{-6}\mathrm{m}^3=1.352\times 10^5$ g/m$^3$.
The molar mass of aam is $m_{\mathrm{mono}}\approx 71$ g/mole. Using the shear modulus of the regular PAAm gels for $C=0.024\%$, the number of monomers between two crosslinkers are approximated as:
\begin{equation}\label{eqn:2}
N_c = \frac{c_{\mathrm{aam}}}{m_{\mathrm{mono}}\mu}N_A = \frac{c_{\mathrm{aam}}}{m_{\mathrm{mono}}\mu/k_B T}N_A\approx 2348
\end{equation}
The mesh size is thus estimated as:
\begin{equation}
    \xi = aN_c^{1/2} = 0.1 \times N_c^{1/2} \approx 5 \mathrm{nm}
\end{equation}
where $a=0.1$ nm is the length of a single bond.

\section{Supplementary Note 2: Thermally activated processes of chain slippage and ruptures}\label{sec:note2}
In experiments, we observed that $G^{1/2}$  for TEAs varies logarithmically with $V_{\mathrm{crack}}$ (Figure 2(a)) if $h_{pen}/h_{dc} \ll 1$, reminiscent of the dynamic adhesion of cell-cell interface\cite{bell1978models}, elastomer\cite{chaudhury1999rate},  and other bonds\cite{pobelov2017dynamic} due to thermally activated bond breaking. To rationalize the results, we adopt a kinetic theory which considers the linkage dissociation as thermally-activated processes\cite{chaudhury1999rate,hui2004failure}. The activation for the dangling chain to slip from the bridging network is assumed to decrease by the applied force (Fig 2f). Using the concept of mechanochemistry, the rate of dissociation of linkage $i$ can be expressed as: 
\begin{equation}\label{eqn:2}
r_i(t) = -\frac{dN_i}{dt} = r_i N_i
\end{equation}
where $i$ can be stitch or slip. $N_i$ is the areal density of linkage $i$, and $r_i$ is the rate constant for linkage dissociation, and is assumed to be dependent on the force applied to the chains $F$ via the Arrhenius law:
\begin{equation}\label{eqn:2}
r_i = r_0\exp{\left(\frac{l_a F}{k_B T}\right)}
\end{equation}
where $r_0 = 1/\tau_{-}$ is the rate constant of the linkage dissociation without adding any force, $\tau_{-}$ is the intrinsic relaxation time of the slip linkage, $k_B T$ is the temperature in the unit of energy, $F$ is the applied force to break a linkage. Note that we have assumed the linkage dissociation process is irreversible, thus the re-association of the linkage is not accounted for. We further consider an individual linkage has a spring constant $k_i$ and is stretched at a fixed velocity $V$ over an averaged bond survival time $\bar{t_i}$. Thus, the averaged force that an individual linkage can bear is given by:$\bar{F_i} = k_i V \bar{t_i}$.

For simplicity, the average linkage survival time $\bar{t_i}$ is estimated by the most probable survival time $t_i^{*}$ corresponding to the maximum of the dissociation rate\cite{evans1997dynamic}: $d^2N/dt^2 = 0$. Therefore, $\bar{t_{i}}$ can be expressed as: 
\begin{equation}\label{eqn:average t}
\bar{t_i}\approx t_i^* = \frac{k_B T}{k_i V l_a}\ln{\left(\frac{l_a k_i V\tau_{-}}{k_B T}\right)}
\end{equation}

The energy released upon breaking linkage $i$ can be expressed as:
\begin{equation}\label{eqn:average e}
e_i = \frac{\bar{F_i}^2}{2k_i} = \frac{V^2\bar{t_i}^2 k_i}{2}
\end{equation}
Substituting Eqn.\ref{eqn:average t} into \ref{eqn:average e} and multiplying $e_i$ with the number density of linkage $N_i$ across an interface yields the energy released by advancing a unit area of an interface held by an array of slip linkages, namely, the adhesion energy:
\begin{equation}\label{eqn:G}
G_i = \left(\frac{N}{2k_i}\right)\left(\frac{k_B T}{l_a}\right)^2 \left[\ln{\left(\frac{l_a k_i V \tau_{-}}{k_B T}\right)}\right]^2 
\end{equation}
We further assume that the crack geometry remains invariant during the steady state peeling process, so $V_{\mathrm{crack}}\sim V$. We express $\tau_{-}=h/(k_B T)\exp\left[E_i/(k_B T)\right]$, where $E_i$  and $h$ are the activation energy of linkage $i$ and Planck constant, respectively.

To calibrate $N_i$, we deduced that it is limited by the area density of the chitosan chains in the bridging network $N_i\approx N_{\mathrm{chi}}$, since its size is presumably larger than that of hydrogel network and the spacing between dangling chains\cite{Yang2018}. If true, Eqn. \ref{eqn:G} suggests that in the absence of stitch linkage, $(G/N_{chi})^{1/2}$ should only depend on $V_{\mathrm{crack}}$. Since we cannot obtain a direct measurement of $N_{chi}$, we adopt an approximation using the areal density of chitosan chains homogeneously dispersed in the solution using the chitosan polymer concentration $c_{\mathrm{chi}}$ (w/v\%) through:
\begin{equation}\label{eqn:N}
N_{chi}\sim\left(\frac{c_{chi}}{M_w}N_a\right)^{2/3}
\end{equation}
Where $M_w\approx 300$kDa, estimated from our GPC test (Fig \ref{figS7}) $N_a = 6\times 10^{23}$ is the Avogadro number. Although Eqn. \ref{eqn:N} provides a rough estimation of $N_{\mathrm{chi}}$, they lead to a reasonable collapse of our data following the reformulation $(G/N_{chi})^{1/2}$ using different values of $c_{chi}$ = 2\%, 1\%, and 0.5\% g/mL (Supplementary Fig 2a and b), confirming the validity to approximate $N_i$ by $N_{chi}$.

Finally, the bridging network may form slip, stitch, or the combination of the two linkages when engaging with the TEA gel, yielding the expression for the total adhesion energy\cite{chaudhury1999rate}:

\begin{equation}\label{eqn:G_final}
    G = \sum_{i} G_i
\end{equation}
with
$$
  G_i = \left(\frac{N_i}{2k_i}\right)\left(\frac{k_B T}{l_a}\right)^2\left[\ln\left(\frac{V_{\mathrm{crack}}k_i l_a h}{(k_B T)^2}\right)+\frac{E_i}{k_B T}\right]^2
$$

Given that an inextensible backing film was attached to the gel, $V_{\mathrm{crack}}$  is determined as the half of the peeling rate in the T-peel test. We then fit the model to the experimental data to estimate $E_i$ and $k_i$.

\section{Supplementary Note 3: Others}\label{sec:note3}
\subsection*{Estimated and fitted values of the effective diffusion coefficients of bridging polymers in the dangling chain layer}

According to the Rouse model, the diffusion coefficient of the polymer in water is given by
\begin{equation}\label{}
D = \frac{k_B T}{N \eta b}
\end{equation}
taking $k_B T= 4.11\times 10^{-21} $J, $N=1000$ the number of units per polymer chain, $\eta = 10^{-3}$ Pa$\times$S, $b=1$ nm the length of a repeating unit of chitosan, it gives $D_{\mathrm{eff}}\approx 5\times 10^{-12}$ m$^2$s$^{-1}$.

\subsection*{Alternative approaches to create slip-linkage topology}
Other approaches exist to create the similar chain-network topology of the slip linkage, such as by placing uncrosslinked polymer chains between hydrogel networks\cite{Yang2018,brown1993effects}. The polymer chains can diffuse into the pre-formed gel networks to form slip entanglement. However, the highly permeable hydrogel would promote the diffusion of polymer chains into the gel matrix, greatly reducing the number of polymer entanglements on the interface.

\subsection*{Summary of topological linkages}
\begin{table}[!ht]
\centering
\caption{{\bf} Types of topological linkages, their constituents (adhesive matrix, targeted surface, and bridging polymer), and the associated data. }
\begin{tabular}{|l|l|l|l|l|}
\hline
\multicolumn{1}{|l|}{\bf Linkage types} & \multicolumn{1}{|l|}{\bf Adhesive matrix} & \multicolumn{1}{|l|}{\bf targeted surface}& \multicolumn{1}{|l|}{\bf Bridging polymer} & \multicolumn{1}{|l|}{\bf Data}\\ \hline
\multirow{2}{4em}{Slip-slip}  & SN TEA PAAm gel& SN TEA PAAm gel & Chitosan & Fig \ref{fig2}d,f, Sup Fig 2a,b \\ 
& $\sim$ & $\sim$ & Gelatin & Fig \ref{fig3}d \\ \hline
Stitch-stitch  & SN regular PAAm gel& SN regular PAAm gel & Chitosan & Fig \ref{fig2}d,f\\ \hline
Slip-stitch  & SN TEA PAAm gel & SN regular PAAm gel & Chitosan & Fig \ref{fig3}c, Sup Fig 2e\\ \hline
\multirow{2}{5em}{Slip-bond}  & SN TEA PAAm gel & VHB elastomer & Chitosan & Fig \ref{fig3}c, Sup Fig 2d\\ 
  & SN TEA PAAm gel & Porcine skin & Chitosan+EDC/NHS & Fig \ref{fig3}e\\ \hline
\multirow{2}{5em}{Sitch-bond}  & SN regular PAAm gel & VHB elastomer & Chitosan & Fig \ref{fig3}c, Sup Fig 2d\\ 
 & SN regular PAAm gel & Porcine skin & Chitosan+EDC/NHS & Fig \ref{fig3}e\\ \hline
\end{tabular}
\label{tab1}
\end{table}

\textbf{Slip or stitch linkages}: To form the slip and stitch linkages, bridging polymer chitosan or gelatin was directly applied to a gel with or without dangling chain layer, and then covered with a permeable adherend.

 \textbf{Bond linkage}: to form bond linkage between bridging chitosan network and tissue surfaces, we utilize the amine groups on chitosan, which can be covalently bonded to the carboxylic acid groups on tissue surfaces with EDC and NHS as coupling reagents\cite{Li2017}: 30 mg of EDC and 30 mg of NHS were added into 1 mL of the chitosan solution for forming covalent bonds with tissue surfaces. To form the bond linkage between bridging polymers and VHB elastomer surfaces, we utilize the Carbonyl bonds on the VHB, which can form imide bonds with the amine group on the chitosan polymer at pH = 4. Besides, the ionic bond formed between NH$^{3+}$ of chitosan of pH $<$ 6.5 and COO$^{-}$ of the VHB of pH $>$ 4.5 can also contribute to the bond linkage. We prepared chitosan of pH 4.5 and hydrogel of pH 7, thus the chitosan can form an interfacial bridging network and can form both type of bonds with VHB surfaces\cite{yang2019design}. 

 \renewcommand\thefigure{S\arabic{figure}} 
 \setcounter{figure}{0}   
\begin{figure}[ht] 
\includegraphics[width=\textwidth]{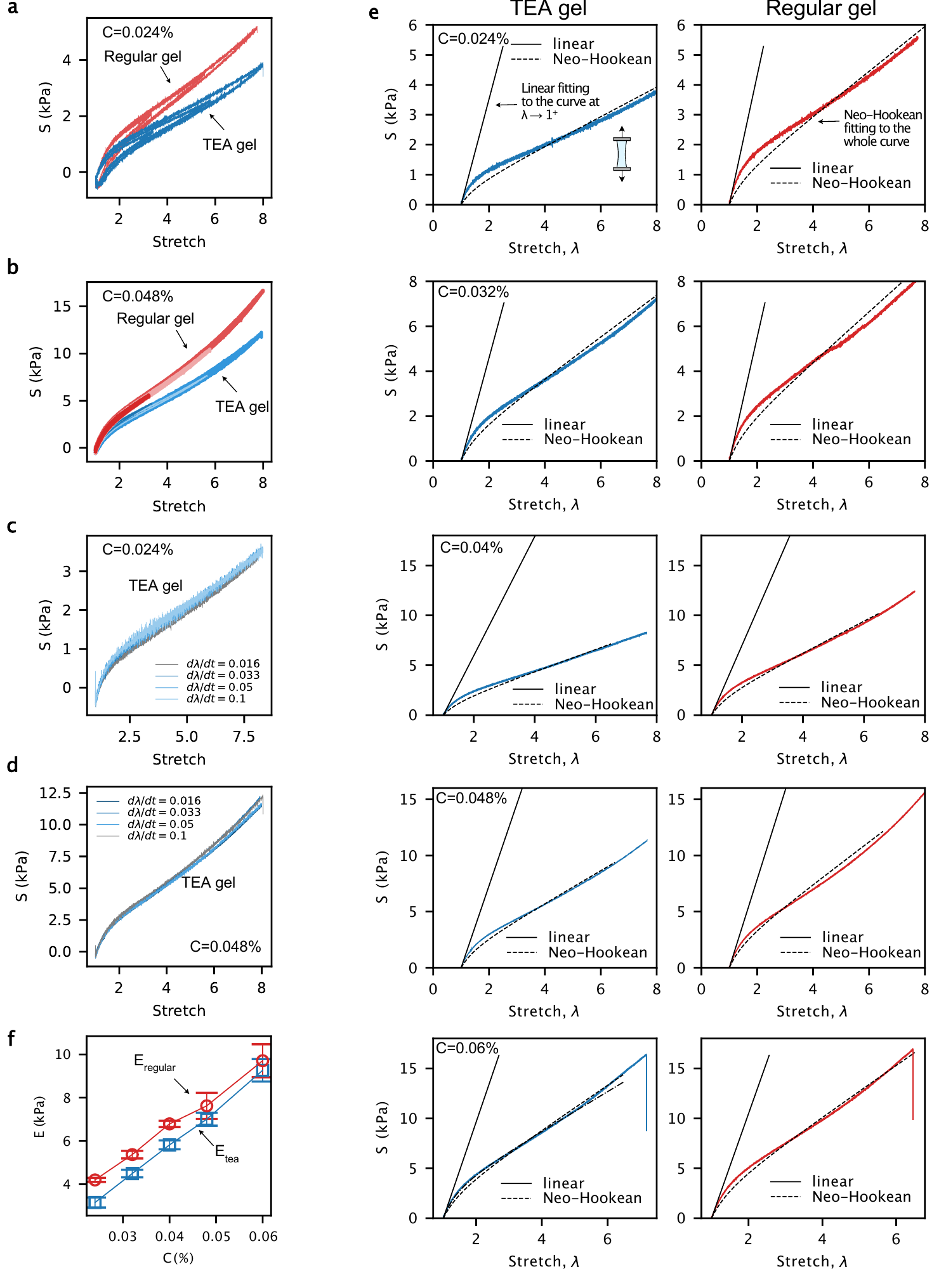}
\caption{
(a) Uniaxial cyclic tensile tests of a TEA and a regular gel made of SN PAAm for $C=0.024\%$. (b) the same curves as (a) for $C=0.048\%$. Uniaxial tensile test results of SN TEA gels with varying strain rates for (c) $C=0.024\%$ and (d) $C=0.048\%$. (e) representative stress-stretch curves of TEA (left) and regular (right) gels for different values of $C$ measured in uniaxial tensile tests. (f) Measured elastic moduli by the linear model for SN TEA and regular gels, plotted as functions of $C$.}
\label{figS1} 
\end{figure}

\begin{figure}[ht] 
    \includegraphics[width=\textwidth]{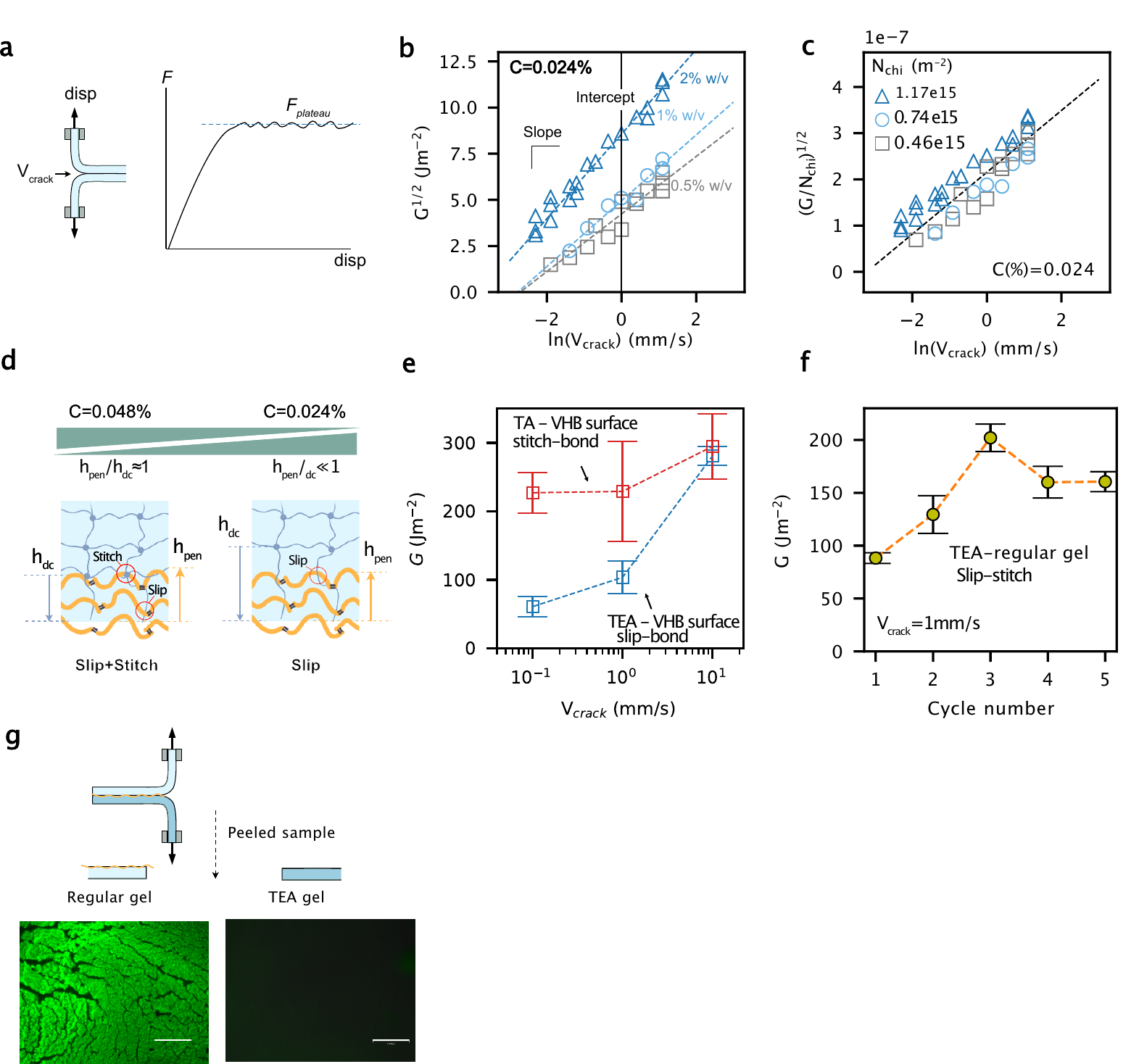}
    \caption{
    (a) Schematic showing the T-peeling test and the resulting force-displacement curves. The adhesion energy is interpret as $G=2F_{plateau}/w$, with $w$ the out-of-plane thickness of the sample. (b) $G^{1/2}$ by the slip-slip linkage plotted as a function of  $\ln V_{\mathrm{crack}}$ for $C=0.024\%$ and varying $c_{chi}$ ($0.5\%, 1\%$, and $2\%$ g/mL) (c) reformulated ($G/N_{chi})^{1/2}$) for $C=0.024\%$, where $N_{chi}$ is estimated using Eqn \ref{eqn:N}. (d) Illustrations showing the effects of entanglement length on the formation of slip and stitch linkages. (e) Stitch- and slip-mediated adhesion on VHB surfaces. (f) Cyclic peeling test on a TEA-regular interface at $V_{\mathrm{crack}}$=1mm/s. After every peeling, a new regular gel adheres to the same TEA gel with newly applied chitosan solution. The adhesion increases and stabilizes after 5 cycles. (g) Peeled regular gel and TEA gel after a peeling test. The interface was bonded with a slip-stitch linkage using fluorescently labeled chitosan. After peeling, the chitosan network remains on the regular gel, but remains minimally in the TEA gel. Scale bars: 1mm.}
    \label{figS2} 
    \end{figure}

    \begin{figure}[ht] 
        \includegraphics[width=\textwidth]{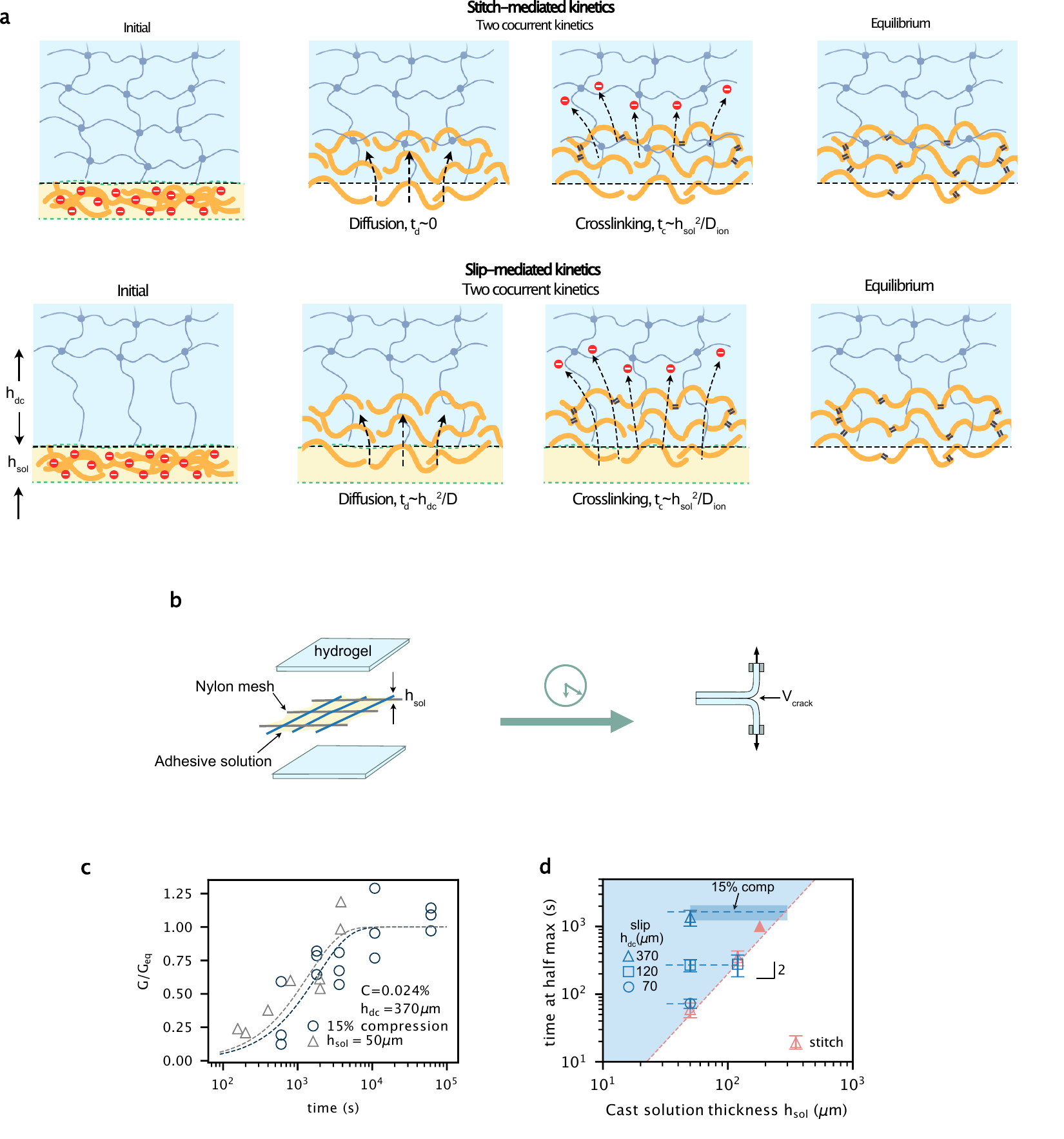}
        \caption{
        (a) Illustrations showing the total kinetics of stitch-mediated (top) and slip-mediated interface (bottom) as two sub kinetic physical processes. (b) Illustration showing the experimental procedure to characterize the adhesion kinetics of TA and TEA. (c) Dimensionless adhesion $G/G_{\mathrm{eq}}$ as functions of waiting time for TEA with $h_{\mathrm{dc}}=370\mu m$. In one case $h_{\mathrm{sol}}=50 \mu m$ was controlled by using nylon mesh of the same thickness at the interface, while in the other case an initial compression strain of $15\%$ was applied to the sample after applying adhesive solution such that $h_{\mathrm{sol}}$ was not well-controlled. No prolonged compression was applied. (d) Summarized $t_{1/2}$ of the slip and stitch-linkage mediated adhesion as functions of $h_{\mathrm{sol}}$ and $h_{\mathrm{dc}}$. Together plotted is the reported $t_{1/2}$ in Steck et al\cite{steck2020topological}, wherein pH-responsive polymer cellulose is used as the bridging polymer (closed markers). Error bars and shaded area represent 95\% confidence interval.}
        \label{figS3}
        \end{figure}

        \begin{figure}[ht] 
            \includegraphics[width=\textwidth]{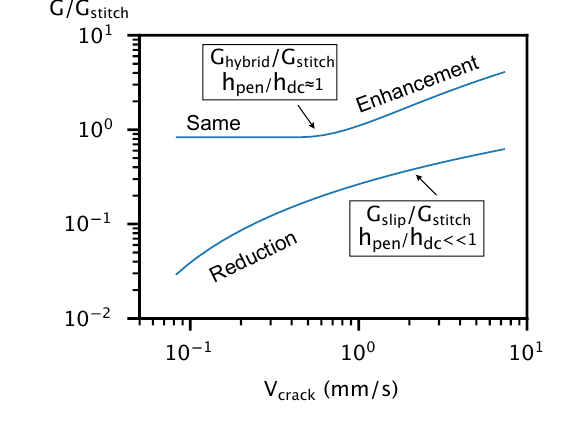}
            \caption{
            Adhesion contrast between the slip and stitch regions in a patterned TEA for given $C$ and $h_{\mathrm{pen}}/h_{\mathrm{dc}}$. The slip linkage becomes the hybrid linkage as $h_{\mathrm{pen}}/h_{\mathrm{dc}}$ gets closer to 1.}
            \label{figS4} 
            \end{figure}

\begin{figure}[ht] 
        \includegraphics[width=\textwidth]{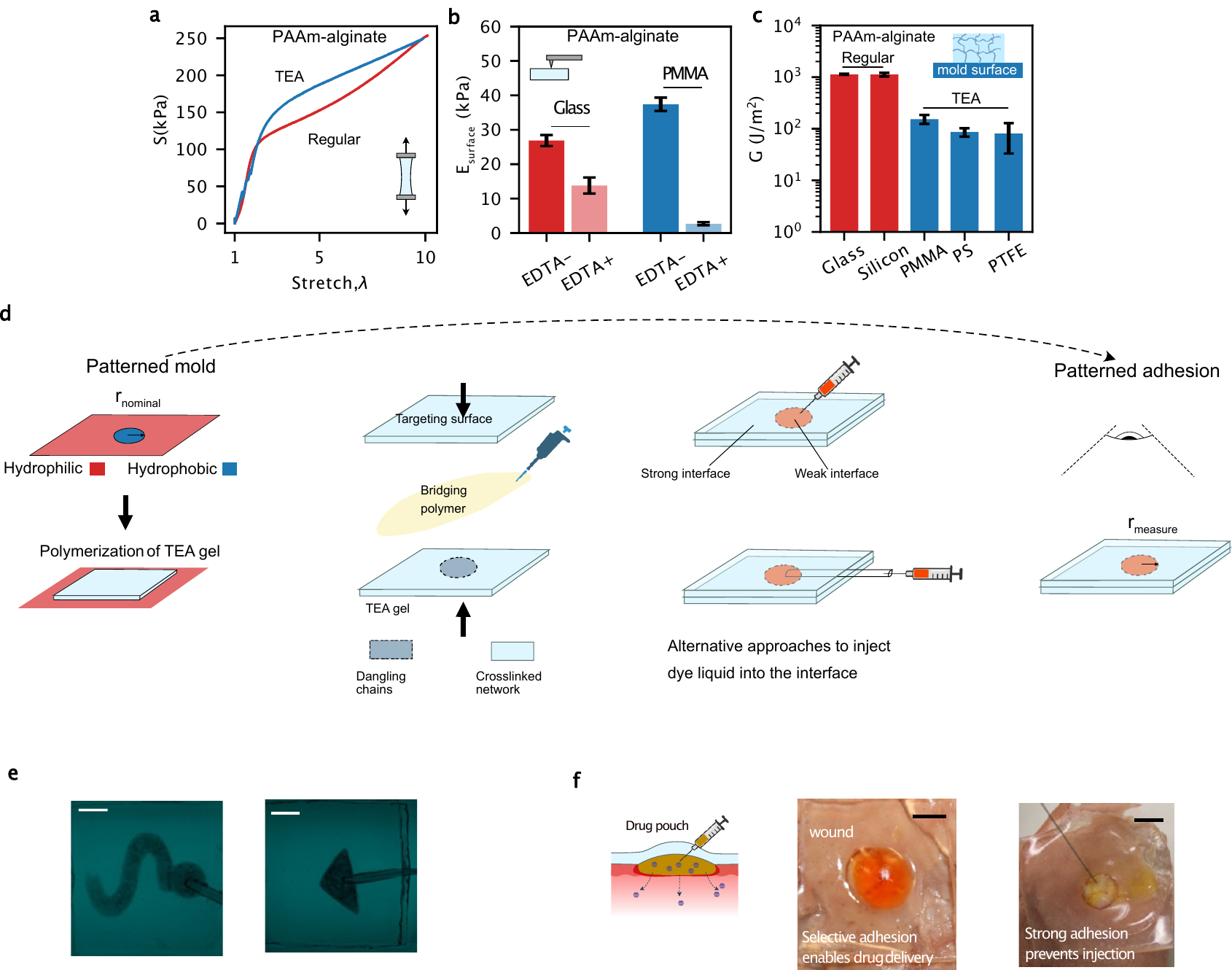}
        \caption{
        (a) Uniaxial tensile test results of a DN TEA and a DN regular PAAm-alginate gels. (b) AFM test on the surface of the DN gels with and without EDTA treatment on the surface. (c) Adhesion of PAAm-alginate gels made on different mold surfaces on porcine skin. (d) Schematic illustrating the procedure for creating spatially programmable adhesion of TEA A needle is used to penetrate the hydrogel into the interface, or a tube was sandwiched between the two gels. Next, liquid dye was injected through the needle or the tube. Finally, a digital camera was used to capture the weak interface through the liquid dye. (e) Patterned interface formed between a TEA PAAm-alginate gel and a regular PAAm-alginate gel. The TEA gel is designed with designed dangling chain regions of different shapes, resulting in weak interfaces of the identical shape. Scale bar: 10 mm. (f) Mock drug is injected to the weak interface of a TEA patch adhered to the wound of a porcine stomach. Instead, if a DN PAAm-alginate TA was adhered to the wounded porcine stomach, strong adhesion to the wound prevents injection of drug.}
        \label{figS5} 
        \end{figure}

\begin{figure}[ht] 
            \includegraphics[width=\textwidth]{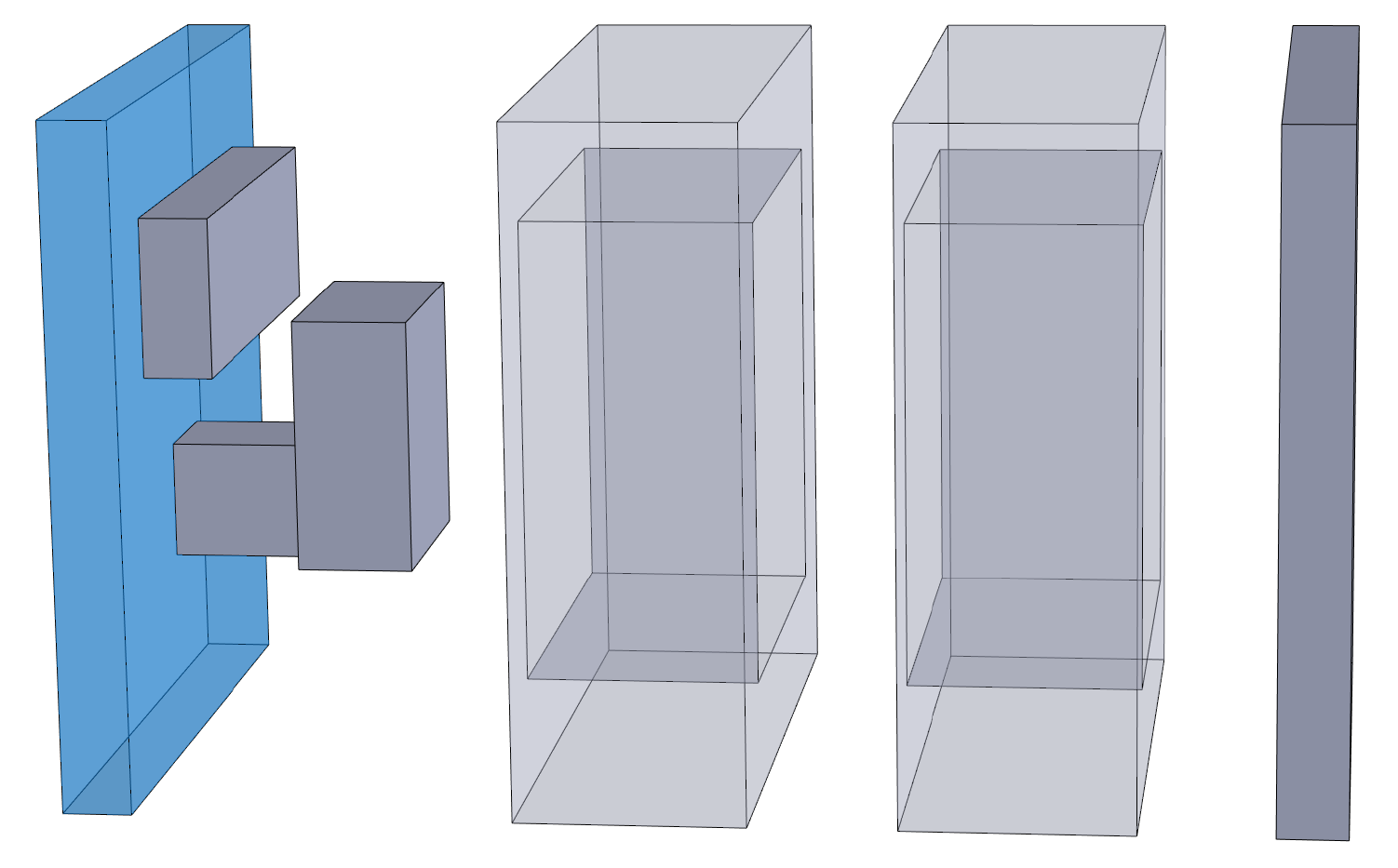}
            \caption{3D printed mold for creating soft actuator modules. The mold parts are 3D-printed using PLA. The cover, indicated in blue, is made of PMMA.}
            \label{figS6} 
            \end{figure}        

\begin{figure}[ht] 
\includegraphics[width=\textwidth]{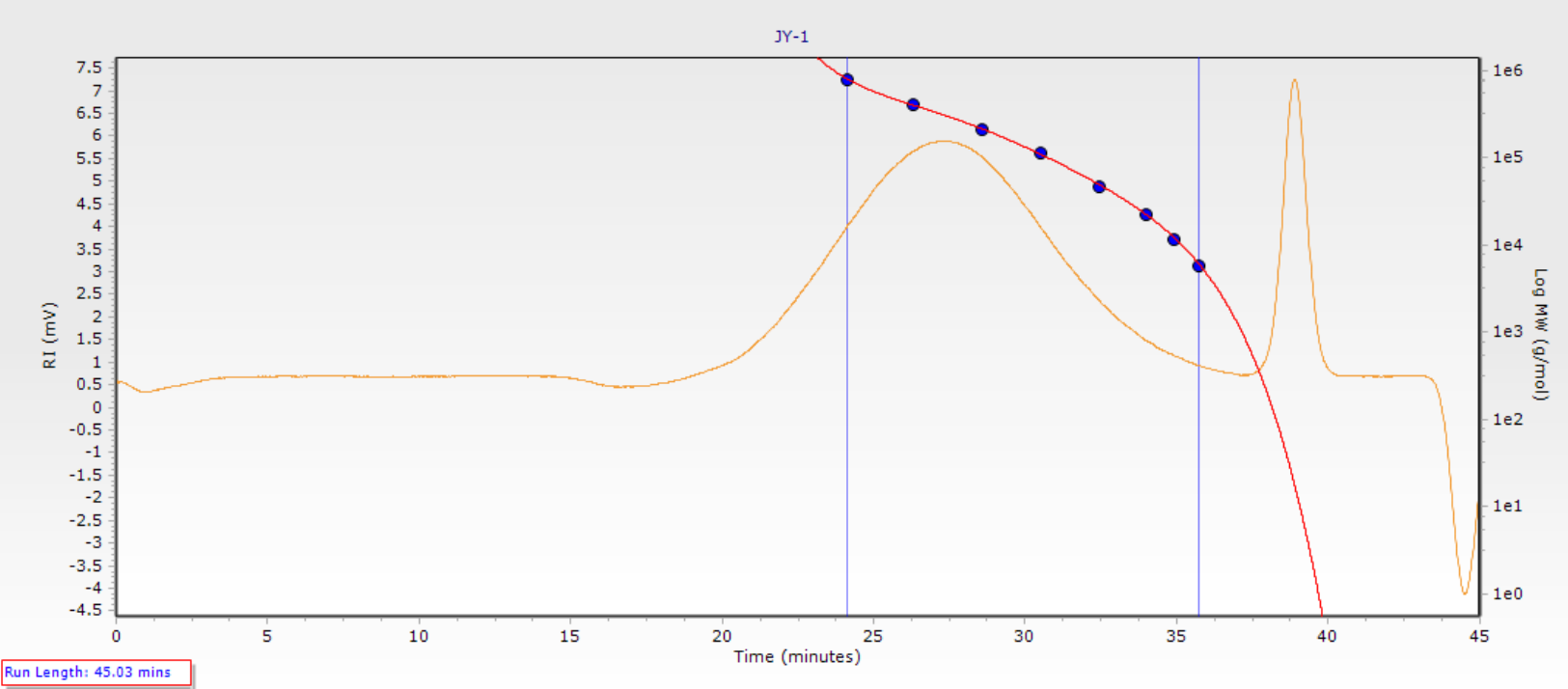}
            \caption{GPC test results.}
            \label{figS7} 
            \end{figure}

\end{document}